\documentclass[fleqn,usenatbib]{mnras}

\newcommand{\NoPrint}[1]{}
\usepackage[multidot]{grffile}  

\usepackage[T1]{fontenc}
\usepackage{ae,aecompl}

\usepackage{graphicx}	
\usepackage{amsmath}	
\usepackage{amssymb}	

\usepackage[normalem]{ulem}

\renewcommand{\arcsec}{{\,arcsec}}
\newcommand{\kms}{{\,km\,s^{-1}}}
\newcommand{\kpc}{{\,kpc}}

\newcommand{\Mpc}{{\,Mpc}}
\newcommand{\h}{{\,h}~}

\newcommand{\etamacro}{\eta^{\rm macro}}
\newcommand{\etasub}{\eta^{\rm sub}}
\newcommand{\RE}{{R_{\rm E}}}
\newcommand{\fsub}{f^{\rm sub}}

\renewcommand{\P}{\mathcal{P}}
\newcommand{\G}{\mathcal{G}}
\newcommand{\Q}{\mathcal{Q}}

\let\vec\mathbfit
\let\mat\mathbfss

\newcommand{\Mmin}{10^{ \rm 6} M_{ \rm \odot}}
\newcommand{\Mmax}{10^{ \rm 10} M_{ \rm \odot}}

\title[Systematic errors in strong gravitational lensing]{Systematic errors in strong gravitational lensing reconstructions, a numerical simulation perspective}

\author[W. Enzi et al.]{Wolfgang Enzi,$^{ \rm 1}$\thanks{E-mail: enzi@mpa-garching.mpg.de}
Simona Vegetti,$^{ \rm 1}$
Giulia Despali,$^{ \rm 1}$ 
Jen-Wei Hsueh,$^{ \rm 2}$
\newauthor
R. Benton Metcalf $^{ \rm 3,4}$
\\
$^{ \rm 1}$Max Planck Institute for Astrophysics, Karl-Schwarzschild-Strasse 1, Garching, Germany\\
$^{ \rm 2}$Kapteyn Astronomical Institute, University of Groningen, P.O. Box 800, 9700AV, Groningen, Netherlands\\
$^{ \rm 3}$Dipartimento di Fisica \& Astronomia, Universita di Bologna, via Gobetti 93/2, 40129 Bologna, Italy\\
$^{ \rm 4}$INAF-Osservatorio Astronomico di Bologna, via Ranzani 1, 40127 Bologna, Italy
}

\date{Accepted 2020 April 26. Received 2020 April 9; in original form 2019 November 6}

\pubyear{2020}

\begin{document}


\label{firstpage}
\pagerange{\pageref{firstpage}--\pageref{lastpage}}
\maketitle

\begin{abstract}

We present the analysis of a sample of twenty-four SLACS-like galaxy-galaxy strong gravitational lens systems with a background source and deflectors from the Illustris-1 simulation. We study the degeneracy between the complex mass distribution of the lenses, substructures, the surface brightness distribution of the sources, and the time delays. Using a novel inference framework based on Approximate Bayesian Computation, we find that for all the considered lens systems, an elliptical and cored power-law mass density distribution provides a good fit to the data. However, the presence of cores in the simulated lenses affects most reconstructions in the form of a Source Position Transformation. The latter leads to a systematic underestimation of the source sizes by 50 per cent on average,  and a fractional error in $H_{0}$ of around $25_{-19}^{+37}$ per cent. 
The analysis of a control sample of twenty-four lens systems, for which we have perfect knowledge about the shape of the lensing potential,
leads to a fractional error on $H_{0}$ of $12_{-3}^{+6}$ per cent. 
We find no degeneracy between complexity in the lensing potential and the inferred amount of substructures.  We recover an average total projected mass fraction in substructures of $f_{\rm sub}<1.7-2.0\times10^{-3}$ at the 68 per cent confidence level in agreement with zero and the fact that all substructures had been removed from the simulation. Our work highlights the need for higher-resolution simulations to quantify the lensing effect of more realistic galactic potentials better, and that additional observational constraint may be required to break existing degeneracies.
\end{abstract}




\section{Introduction}

Galaxy-galaxy strong gravitational lensing is a powerful tool to investigate a large number of diverse astrophysical and cosmological inquiries \citep[see][and references therein]{treu2010}. For example, in the detailed analysis of the physical and kinematical properties of distant galaxies strong lensing provides the magnification that allows one to overcome the observational limitations of low signal-to-noise ratio and spatial resolution 
\citep[e.g.][]{shirazi2014,rybak2015,rizzo2018,spingola2019}. Measurements of the time-delay between the multiple lensed images of time-varying sources have been demonstrated to provide a promising constraint on the Hubble constant $H_0$ and weak constraints on other cosmological parameters  \citep[e.g.][]{suyu2010,wong2019,chen2019}.
The sensitivity of lensing to total mass has been used to quantify the amount of low-mass dark matter haloes and thereby the properties of dark matter, both with lensed quasars \citep[][]{mao1998,nierenberg2014,gilman2019,hsueh2019} and lensed galaxies \citep[][]{koopmans2005,vegetti2009,vegetti2012,hezaveh2016,vegetti2010b}.

In order for these studies to be robust, a good understanding of the lenses gravitational potential is essential.
For example,  \citet{ritondale2019b} have shown from the analysis of the BELLS GALLERY samples, that in certain cases it is necessary to extend the lensing mass density distribution beyond the standard assumption of a single or multiple power-laws in order to obtain a correct focusing of the reconstructed background source.
Using hydrodynamical simulations, \citet{xu2017} have shown that power-law mass models can bias the inference on the Hubble constant by up to 50 per cent, a deviation that is related to the mass sheet degeneracy \citep{falco1985}. 
Using both numerical simulations and mock observations \citet{gilman2018} and \citet{hsueh2018} have demonstrated that complex baryonic structures in elliptical galaxies can contribute to 8-10 percent of the strength of flux-ratio anomalies, and therefore constitute a potential bias in the inferred properties of dark matter \citep[see also][]{moeller2003,xu2015}.
Similarly,  \citet{hsueh2016,hsueh2017} have shown that lens galaxies with a significant edge-on disc are characterised by strong flux-ratio anomalies that can be explained by the presence of the disc without the need to include a significant population of substructures.
Recently, from the analysis of extended lensed images from the the BELLS GALLERY sample, \citet{ritondale2019b} have concluded that complex mass distributions can emulate the effects of substructures and, therefore, lead to false-positive detections.
More generally, \citet{unruh2017} have shown that source position transformations \citep{schneider2014} can explain why independent studies of the same lens system may result in different inferred lens parameters.

In this work, we quantify the systematic errors that may be induced by departures from simple power-lass lensing mass distribution on the different lensing observables using galaxies taken from the cosmological hydrodynamical simulation Illustris-1 \citep{vogelsberger2014}.
We remove all substructures from the simulations and create mock lensing observations emulating the SLACS \citep{Bolton06} survey.
We then model these data using a novel inference approach based on Approximate Bayesian Computation (ABC), under the classical assumption of a cored power-law mass model.
We then infer the fraction of projected mass contained in substructures $\fsub$ to quantify the degeneracy between substructures, time delays (which are relevant in the determination of $H_{0}$), other forms of complexity in the lensing potential, and the source surface brightness distribution.
The paper is structured as follows: in Section \ref{sec:simulations} we give a short overview of the Illustris-1 simulations and specify how we create mock data. In Section \ref{sec:lens_modelling} we describe the physical model as well as the statistical method that we use for the reconstruction of the mock data. In Section \ref{sec:results} we present and discuss our results. In Section \ref{sec:summary} we conclude our study by summarising our main findings.

\begin{figure*} 
\includegraphics[width=\textwidth]{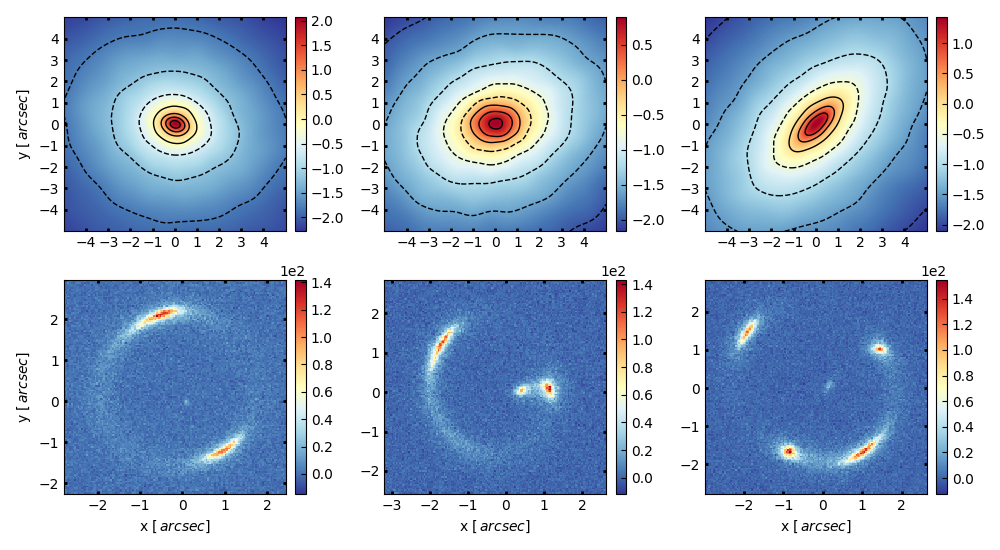} 

\caption{Top panels: the convergence maps of three representative galaxies selected from the Illustris-1 simulation. All substructures have been removed in these images. The dashed lines represent iso-convergence contours. Bottom panels: the corresponding mock lensing observations in our sample of lens systemss. 
} 
\label{fig:Illustris1_kappa} 
\end{figure*}


\section{Lensing data generation}\label{sec:simulations}

In order to study the lensing signal from galaxies with complex mass distributions, we focus on simulated galaxies from the hydro-dynamical numerical simulation Illustris-1. In this section, we provide a short description of the simulation and how we generated mock lensing observations from it.

\subsection{Lens and source galaxies selection}
\label{mock_creation}

The Illustris-1 project is a series of hydro-dynamical numerical simulations of cosmological volumes that follow the evolution of dark matter, cosmic gas, stars, and supermassive black holes from a starting redshift of $z  = 127$  to the present time, all while accounting for realistic baryonic physics. 
In this work, we use the main run of Illustris-1, which has a box size of $106.5 \Mpc$  in all three dimensions and contains $1820^{ \rm 3}$  dark matter particles and (initial) gas cells. The simulations were run using the  moving-mesh   {\sc Arepo}  code \citep{springel10}. 
The dark matter particle mass is $ 6.3 \times10^{ \rm 6}M_{ \rm \odot}$ and the initial gas particle  mass is 1.3 $\times10^{ \rm 6}M_{ \rm \odot}$, while the  simulation uses softening lengths for dark matter and baryons that are $\epsilon_{\rm dm}$=1.4 kpc and $\epsilon_{\rm b}$=0.7 kpc, respectively.
The simulations adopt the cosmological parameters $\Omega_m = 0.2726$, $\Omega_\Lambda = 0.7274$, $\Omega_b = 0.0456$, $\sigma_8 = 0.809$, $n_s = 0.963,$ and $H_0 = 100 \h\times \kms/\Mpc$ with $\h = 0.704$, which are consistent with the latest Wilkinson Microwave Anisotropy Probe 9 measurements \citep{hinshaw2013}.

From the catalogue of Illustris-1 we choose ten galaxies at redshift $z= 0.2$ to be our sample of lenses. 
These are analogues of the SLACS lenses \citep{Bolton06} and were selected by \citet{despali2017} based on a number of properties:
the total halo mass, stellar mass of the central galaxy, stellar effective radius, velocity dispersion, and  dynamical properties compatible with an early-type morphology. 
In particular, they selected galaxies according to the fraction of stellar mass showing specific circularities $\epsilon  = J_{\rm z} \big / J(E)$, where $J_{\rm z}$ is the angular momentum of an individual stellar particle and $J(E)$ is the maximum angular momentum within its local environment.
The fractions $f_{\epsilon <0.0}$ and $f_{\epsilon >0.7}$ are indicators of an early-type morphology \citep[see e.g.][]{genel2015,teklu15}, typical for lens galaxies at the considered redshift.

This selection ensures that we avoid galaxies with a significant disc component, for which the lensing properties have already been investigated by \citet{ hsueh2018}. In Table \ref{tab:namesinfo} we list the main properties of the considered sample of lens galaxies.
We also select a compact  Illustris-1 galaxy at redshift $z=0.6$ as our background source to be lensed. We use the same source galaxy for all created mocks.

\subsection{Ray tracing}\label{glamer}

For each of the ten lens galaxies in our sample, we consider the projection of their mass density distribution onto the three principal planes and use the ray-tracing code GLAMER \citep{metcalf2014,petkova2014}\footnote{\href{http://glenco.github.io/glamer/}{http://glenco.github.io/glamer/}} to generate $3\times 10$ maps of the lensed source surface brightness distribution. 
The focus of this paper is on the lensing effect of small scale complexity in the lens mass distribution such as dense baryonic structures from tidally disrupted satellites and general departures from simple single power-law models.   Therefore, we have removed all particles belonging to dark matter substructures and only considered the contribution from the main halo. However, it is worth mentioning that in the Illustris-1 simulation, the minimum resolved substructure mass is of the order of $\simeq 10^{8}M_{\odot}$. For each main halo we take all its particles within $R_{200} \approx R_{\rm vir}$ as identified by the SUBFIND algorithm \citep{springel2001}. 

For a given collection of particles (i.e. their positions, masses,  and smoothing lengths as  used in the  original simulation), GLAMER determines the corresponding deflection angle maps with an efficient tree algorithm.
In order to avoid unrealistic deflections of the light rays, each particle is represented by a B-spline in three dimensions as it is commonly done in smooth particle hydrodynamics (SPH) simulations. 
We set the size of the smoothing length to the distance of the $N^{ \rm smooth}$-th nearest neighbour. We choose $N^{\rm smooth}=64$, which corresponds to scales $\lesssimÂ \mathcal{O}(\kpc)$.
This scheme provides smaller smoothing lengths where the particles are dense (for example at the centres of haloes), and larger ones where the particles are sparse and shot noise would otherwise pose a problem.
We follow a similar approach for the baryonic cells by treating them as particles with a corresponding smoothing length in order to make them compatible with the implementation of GLAMER. The minimum and average smoothing lengths over the whole sample are $  0.6-0.8 \kpc$  and $ 16 \kpc$ respectively, which also taking into account the softening length of the simulation tell us that we can safely resolve inhomogeneities on  scales $> \mathcal{O}(\kpc)$. Notice that the pixel size of the mock images is $0.04 \arcsec$.  

As an example, Figure \ref{fig:Illustris1_kappa} shows the convergence maps (i.e. surface mass density normalised by the critical surface mass density) of three representative Illustris-1 galaxies. The presence of small scale features in the mass distribution, which are typically not captured by simplified parametric profiles, are highlighted by the irregular iso-convergence contours.

The galaxies in the Illustris-1 sample show unrealistically large cores.
This effect is a well-known challenge for simulations and arises from finite smoothing lengths. 
Depending on the lensing configuration, these large cores lead to lensed images with morphologies (see middle-bottom panel of Fig. \ref{fig:Illustris1_kappa} for an example) which are rarely observed in real gravitational lens systems, where the central mass density distribution may be cuspier.
For this reason we remove six realisations from our set and in the following we will only focus on the 24 systems with realistic lensed images. 
A similar problem was recently identified by \cite{mukherjee2018} in a study of the {\sc Eagle} simulation \citep{schaye2015}.

\subsection{Observational effects}\label{sec:obs_eff}

In order to generate mock HST observations we convolve the lensed surface brightness distribution $I(\vec x)$ with the point-spread function (PSF) $p(\vec x)$ from the WFC3 camera in the F606W filter and add the contribution of observational noise. In particular, for each of the lensed images in our sample, we draw a realisation of Gaussian random noise $\vec n(\vec x)$, with
\begin{equation}
P(\vec n | \mat N ) = \G (\vec n, \mat N):= \exp \Big(-\frac{1}{2} \vec n^t  \mat N^{-1} \vec n  \Big)/ \sqrt{\det(2 \pi  \mat N)} \,,
\end{equation}
which includes the contribution of a constant background $b$ and another term proportional to the signal, which approximates the Poisson noise of photon counts. We assume the noise to be uncorrelated and therefore entries of the covariance matrix are multiplied by the Kronecker delta $\delta_{ \rm ij}^K$:
\begin{equation}
\mat N_{ \rm ij} =   \Big[ b +  (p * I )(\vec x_i)  \Big] \times  \delta_{ \rm ij}^K\,.
\end{equation}
The level of noise is chosen in such a way to match the signal-to-noise ratio of the SLACS lenses analysed by \citet{vegetti2014}.

\begin{table*}
\caption{The properties of the lens galaxies selected from the Illustris-1 simulation. From left to right,  we list the virial mass $M_{ \rm vir}$, the mass of the dark matter component $M_{ \rm dm}$, the stellar mass $M_{ \rm *}$, the gas mass $M_{ \rm gas}$, the fraction of stellar mass showing a circularity higher than 0.7 $f_{ \rm \epsilon>0.7}$ and lower than 0.0  $f_{ \rm \epsilon<0.0}$, as well as the virial radius $r_{ \rm vir}$ and stellar radius $r_{ \rm *}$. }
\label{tab:namesinfo}

\begin{tabular}{ |c|c|c|c|c|c|c|c|c|c|c|c|c|c| }

\hline

$ID_{ \rm fof}$  &$M_{ \rm vir} $ &$M_{ \rm dm} $ &$ M_{ \rm *} $ &$M_{ \rm gas} $  &$f_{ \rm \epsilon>0.7}$ &$f_{ \rm \epsilon<0.0}$ &$r_{ \rm vir} $ &$ r_{ \rm *} $ \\   
&$M_\odot / \h$ &$M_\odot / \h$ &$ M_\odot /\h$ &$M_\odot / \h$  &$1$ &$1$ &$\kpc /\h$ &$\kpc /\h$ 
\\\hline
28  & 2.314e+13 & 1.910e+13 & 4.107e+11 & 1.582e+11 & 1.580e-01 & 6.522e-01 & 6.319e+02 & 1.025e+01 \\  40 &   2.789e+13 & 2.404e+13 & 5.945e+11 & 1.396e+11 & 1.662e-01 & 6.563e-01 & 6.725e+02 & 1.458e+01 \\  51  & 2.094e+13 & 1.906e+13 & 4.691e+11 & 8.335e+10 & 7.534e-02 & 9.388e-01 & 6.112e+02 & 1.205e+01 \\  55  & 2.327e+13 & 1.953e+13 & 4.341e+11 & 2.664e+11 & 1.323e-01 & 7.430e-01 & 6.330e+02 & 1.539e+01 \\  65 & 1.749e+13 & 1.543e+13 & 6.486e+11 & 3.023e+10 & 1.284e-01 & 8.440e-01 & 5.756e+02 & 1.916e+01 \\  84  & 1.179e+13 & 1.134e+13 & 3.427e+11 & 5.861e+10 & 1.664e-01 & 7.242e-01 & 5.047e+02 & 1.411e+01 \\  91 &  1.355e+13 & 1.085e+13 & 2.147e+11 & 2.446e+10 & 6.730e-02 & 8.781e-01 & 5.286e+02 & 8.816e+00 \\  95  & 1.094e+13 & 9.443e+12 & 3.164e+11 & 1.371e+11 & 5.936e-02 & 9.282e-01 & 4.922e+02 & 1.225e+01 \\  121 & 6.297e+12 & 5.658e+12 & 1.389e+11 & 1.550e+11 & 1.085e-01 & 7.531e-01 & 4.095e+02 & 2.656e+00 \\  140  & 8.737e+12 & 7.818e+12 & 2.268e+11 & 4.465e+10 & 1.077e-01 & 7.056e-01 & 4.567e+02 & 7.336e+00 \\  
\hline
\end{tabular}

\end{table*}

\section{Lens modelling}
\label{sec:lens_modelling}
In this section, we discuss the physical components that define our inference problem and present a novel inference approach based on Approximate Bayesian Computation (ABC). For an overview on ABC in cosmology, we refer the reader to \citet{akaret2015} and to \citet{birrer2017}, and \citet{gilman2018,gilman2019} for examples of applications in the context of strong gravitational lensing.

\subsection{Physical model}

The physical model includes the following unknown components: the surface brightness distribution of the source galaxy, the  mass distribution of the lens galaxy and the amount of lens-galaxy mass contained in substructures.
\subsubsection{Source Surface Brightness distribution} \label{sec:source}

We follow 
 \cite{vegetti2009_2} 
and define the source surface brightness distribution $s(\vec x)$ on an adaptive grid, so that $\vec s_i=s(\vec x_i)$ corresponds to the brightness value at the position $\vec x_i$ of the $i$-th vertex of a Delaunay mesh. 
The mesh vertices are obtained by ray tracing a subset of the pixels of the observed data $d(\vec x)$ to the source plane. 
The source surface brightness at each vertex constitutes a free nuisance parameter of the model, while the source brightness within the triangles spanned by these vertices is determined via linear interpolation.

\subsubsection{Lens projected mass distribution}\label{Macromodel}

We parametrise the projected mass distribution of the main deflector with an elliptical cored power-law profile of convergence
\begin{equation}
\kappa(\rho) = \frac{\kappa_0 \left(2 - \frac{\gamma}{2} \right)\,q^{-1/2}}{2( r_c^2 +\rho^2)^{(\gamma-1)/2}}\,, 
\end{equation}
where $\kappa_0$ is the amplitude, $\gamma$ the 3D slope, $q$ the projected minor-to-major axis ratio in projection, $\theta$ the angular orientation of the major axis (defined North to East), $x_0$ and $y_0$ the  centre position coordinates, and $r_c$ the core radius.  After a rotation by $\theta$ and a translation of $(x_0, y_0)$  the ellipse radius $\rho$ is defined via $\rho^2  = x^2  + y^2 /  q^2$.
 Together with the strength $\Gamma$ and position angle $\Gamma_\theta$ of an external shear component, these geometrical parameters constitute part of the key target parameters in the reconstruction process and we collectively refer to them as macro-model parameters or $\etamacro = (\kappa_0,q,\theta,x_0,y_0,r_c,\gamma,\Gamma,\Gamma_\theta)$. We use the {FASTELL}\footnote{\href{http://wise-obs.tau.ac.il/~barkana/codes.html}{http://wise-obs.tau.ac.il/{$\sim$}barkana/codes.html}} library developed by \citet{barkana1998} to calculate deflection angles of non-isothermal ($\gamma \neq 2$) cored power laws.

\subsubsection{Substructure}
\label{sec:substructure_model}

Each substructure is assumed to have a spherical NFW mass profile \citep{navarro1996} with a concentration given by the mass-concentration relation of \citet{duffy2008} and a virial radius according to \citet{bullock2001}.
We do not consider the effects of tidal truncation nor variations in the concentration-mass relation as a function of the distance from the host centre, as these are of secondary importance \citep[see, e.g.][]{despali2018}.
We assume the number density of substructure in mass bins of width $[m,m+dm]$ (i.e. the substructure mass function) to be well described by a power-law \citep{Springel2008}
\begin{equation}
\label{equ:mf}
\frac{ d n^{ \rm sub}(m) }{d m}  = n_0 \times m^{ \rm -\alpha} \,,
\end{equation}
with a slope of $\alpha \approx 1.9$ and a normalisation $n_0$ which depends in general on the redshift and the mass of the host galaxy \citep{gao2011,xu2015}.
Here, we neglect both dependencies because the lens galaxies are all at the same redshift and span a narrow range in virial mass.
The amplitude of the mass function is determined by the fraction of projected total mass of the host galaxy within the Einstein radius $\RE$ (calculated according to Appendix \ref{sec:einstein_radius}) contained in substructure, $\fsub(<\RE)$ (see Appendix \ref{sec:mass_function_amplitude} for a derivation).
The number, masses and projected positions of each substructure realisation $\etasub$ are a free nuisance parameter, while $\fsub(<\RE)$ is the main target parameter of the inference problem.
In the following, we drop the argument of $\fsub(<\RE)$ and write $\fsub$, while always referring to the fraction within the Einstein radius unless stated otherwise.

\begin{figure*}
    \centering
    \includegraphics[width=\textwidth]{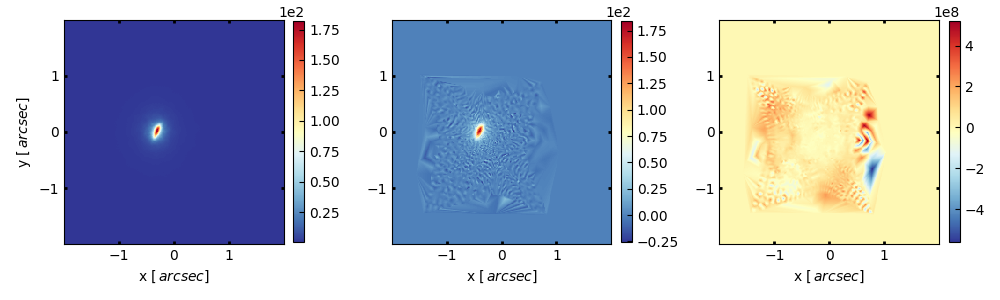}

    \caption{From left to right: the input source selected from the Illustris-1 simulation, the source inferred from the lens modelling of one of the lens systems in our sample, a prior sample of the surface brightness distribution. Notice that in the latter case the resulting field realization does not resemble a realistic source brightness distribution. The middle and right plots were generated using the same regularisation form $\mat S$ and  strength $\lambda$. 
    }
    \label{fig:source_grf}
\end{figure*}

\subsection{Reconstruction approach}
\label{sec:reconstruction}

In this section, we describe the statistical method we use to infer the macro-model parameters $\eta^{\rm macro}$ and the substructure mass fraction $\fsub$ while marginalising over the source surface brightness distribution and the substructure realisations.

\subsubsection{Priors}
\label{sec:prior}

For the parameters describing the lensing mass distribution we choose the following priors: a uniform prior for the amplitude $\kappa_0$, the axis ratio $q$, the angular orientation $\theta$ and the slope $\gamma$, a Gaussian prior for the position $x_0,y_0$ and a log normal distribution for the core radius $r_c$. 

The tessellated source surface brightness distribution is drawn from a Gaussian regularising prior:
\begin{equation}
\P(\vec s| \mat S, \lambda) = \G(\vec s,\lambda^{-1} \mat S)\,,
\end{equation}
where the matrix $\mat S$ determines the form of the regularisation and $\lambda$ its strength. In particular, we choose a form of $\mat S$ that penalises either the gradient or the curvature at the vertices spanning the mesh \citep[e.g.][]{suyu2006}. We choose a log uniform prior on the regularisation strength  \citep[e.g.][]{vegetti2009_2}.

The probability of a substructure having a mass $m$ in $[m,m+dm]$ is related to the mass function introduced in Section \ref{sec:substructure_model} via the following relation:
\begin{equation}
\P(m) = \frac{ d n^{ \rm sub}}{ d m} \Bigg / \int_{ \rm \Mmin}^{ \rm \Mmax} d m  \frac{d n^{ \rm sub}}{dm} \,.
\end{equation}
We assume that the number of substructure $N^{ \rm sub}$ follows a Poisson distribution 
\begin{equation}
\P(N^{ \rm sub}|\mu) =\frac{ e^{ \rm -\mu} \mu^{ N^{ \rm sub}}}{N^{ \rm sub} !}\,,
\end{equation}
of mean value $\mu$ given by
\begin{align}
 \mu = A^{\rm data} \times \frac{\fsub(<\RE)  \times \Sigma_{ \rm crit}  }{ <m>_{ \rm \P(m) }} \,,
\end{align}
where $ A^{\rm data}$ is the area covered by the data (see Appendix \ref{sec:mass_function_amplitude} for a derivation).
We assume a prior probability for the substructure projected  positions which is uniform within the extent of the considered lens plane \citep{xu2015}.

We choose a prior of $\P(\fsub) \propto \big(\fsub\big)^{-1/2}$, which is the Jeffreys prior of the likelihood $\P(N^{\rm sub} | \fsub)$ (see Appendix \ref{sec:jeffreys_prior}).

\subsubsection{Macro-model reconstruction} \label{sec:macro_rec}

In order to find the optimal macro-model parameters, we propose a modified version of Approximate Bayesian Computation (ABC). ABC is usually motivated by inference problems, for which the Likelihood is not accessible. In Appendix \ref{sec:ABC_append}, we discuss why we follow this approach in this work and why the evaluation of the Likelihood poses a challenge for our analysis.
In its simplest form ABC is a rejection sampling algorithm in which prior samples of the target parameters -  in this case samples of $\eta^{\rm macro}$ - are accepted/rejected according to the distance between the  model realisation  they generate $\vec b(\vec x)$ and the observed data $\vec d(\vec x)$.
Accepting the subset of samples closest to the real observation yields a set of samples approximating the posterior distribution. Accepting only perfect matches would yield the exact posterior distribution, although these are extremely unlikely for a finite number of realisations.
Therefore, it is customary to set a distance threshold or accept only the fraction of the drawn samples that are closest to the data.
Alternatively, one can compare statistical summaries $\Phi(\vec b(\vec x, \eta^{\rm macro}_i))$ and $\Phi(\vec d)$ rather than  the mock data realisations and observations directly to each other. 
This approach improves the acceptance rate of the procedure, but only approximates the true posterior. As described below, the latter is the approach adopted in this paper.

A fully forward modelling method requires that we draw  a realisation of the source surface brightness  distribution from a chosen prior (see Section \ref{sec:prior}). 
This requirement, together with the pixelated nature of our source, poses a challenge as every sampled source brightness distribution is almost guaranteed to be far away from the true one as illustrated in Figure \ref{fig:source_grf}.
Searching through the parameter space of the target parameters $\eta^{\rm macro}$ while drawing new source realisations in every sample would, therefore, require a large number of samples, rendering the method infeasible. 
For this reason, we would like to obtain a distance measure that directly compares a lens configuration to the observations, without having to draw a source realisation explicitly.
We propose, therefore, to use the following Evidence as a proxy to a conventional distance measure:
\begin{multline}
 dist(\vec d, \etamacro)  \approx \\
 \approx - 2 \log \P(\vec d |  \mat  L , \lambda)   = \\ =
- 2 \log  \Big[ \int d \vec s ~d \vec n~ \P(\vec d  | \mat  L, \vec s, \vec n)  \P(\vec s| \mat S, \lambda) \P(\vec n | \mat N) \big] =\\  = -2 \log \big[ \G(\vec d, \mat N +  \lambda^{-1} \mat P\mat L\mat S \mat L^T\mat P^T) \Big]\,, 
\label{equ:evidence}
\end{multline}
where $\mat L = \mat L (\etamacro)$ is the lensing operator for a macro-model $\etamacro$ and $\mat P$ a blurring matrix reproducing the effects of the point spread function $p(\vec x)$.  
In particular, since one can evaluate the integral analytically, it will also include the more realistic source configurations that are of interest in the inference. Sampling individual source realisations will, in contrast, tend to be far away from what is expected for real galaxies (as discussed in the previous section).
Since the priors on the source and the noise are Gaussian distributions, the maximum-posterior source appears implicitly in the exponent of the Posterior probability distribution.

Heuristically, we expect those lens configurations with higher Evidence to match the data more frequently than those with lower Evidence. 
Another, more Bayesian viewpoint is that the Evidence can be interpreted as a measure of credibility of the data $\vec d$ on the hypothesis of a $\eta^{\rm{macro}}_i$ \citep[see e.g.][]{cox1946}.
Both arguments indicate that we should prefer to accept higher Evidence samples to obtain a posterior approximation.
Following this consideration, we rank the drawn samples in order of their Evidence values and accept the fraction of samples with the highest values. In order to improve the convergence behaviour further, we introduce a proposal distribution $\Q(\eta^{\rm{macro}})$, which we recursively update until it generates samples close to the posterior distribution. 

Our inference on the macro-model parameters then proceeds in an iterative fashion as follows:\\

\noindent (i) We draw $N^{ \rm samp}$ samples of macro-model parameters $\eta^{\rm macro}_i$ from the proposal distribution. At the first iteration we initialise the proposal distribution to match the prior introduced in Section \ref{sec:prior}, i.e. $\Q(\eta^{\rm{macro}}) = \P(\eta^{\rm{macro}})$.\\

\noindent (ii) For each sample we calculate its importance sampling weight $w_i =\P(\etamacro_i) /\Q(\etamacro_i)$ and Evidence $ \P_i = \P(\vec d| L( \etamacro_i) )$. We accept the $N^{\rm acc}$ samples with the highest values of $P_i\times w_i$ (see Appendix \ref{sec:importance_acceptance} for further details). 
Whenever we find that the effective number of samples 
\begin{equation}
N^{ \rm eff} =   \Bigg( \sum_{ \rm i=1}^{ N^{\rm samp}} \P_i \times w_i \Bigg)^2 \Bigg/ \Bigg( \sum_{ \rm i=1}^{ N^{\rm samp}} \P_i^2 \times w_i^2 \Bigg)
\end{equation}
is higher than $N^{\rm acc}$ we accept this effective number of samples instead. 
A chosen value of $N^{\rm acc}$ therefore determines the minimum relative acceptance threshold of $\epsilon = \frac{N^{\rm acc}}{N^{\rm samp}}$.
The above acceptance strategy ensures that we do not underestimate the region within which we have to probe the parameter space in the following iteration.\\

\noindent (iii) We update the proposal distribution according to the following expression
\begin{equation}
\Q(\eta^{{\rm macro}}) =q \times  \G(\eta^{{\rm macro}} - \mu, \Sigma) + (1 - q) \times \P(\eta^{{\rm macro}})\,,
\end{equation}
where $\G(\etamacro-\mu, \Sigma)$ is a Gaussian distribution estimated from the accepted samples. Above, $q$ determines the fraction of samples drawn from the prior versus the fraction drawn from the estimated Gaussian distribution. We set $q=0.25$, so that one fourth of all drawn samples are used to search through the initial parameter space thereby reducing the probability that we converge to a local minimum. 

We repeat the steps (i) to (iii) $N^{ \rm step}$ times, which results in a large number of samples, most of them drawn in the interesting regions of the parameter space. Following the discussion in Appendix \ref{sec:importance_acceptance} the approximate posterior is  given by the distribution of the accepted parameters:
\begin{equation}
\P^{\rm ABC}( \eta^{{\rm macro}} | \vec d) = \frac{1}{K^{\rm acc}} \sum_{\rm i \in {\rm acc}} \delta^D ( \etamacro - \etamacro_i)\,,
\label{equ:abc_post}
\end{equation}
with $K^{\rm acc}$ being the total number of accepted samples so far.  The expectation value of some function $f(\etamacro)$  is then given by the average of the accepted samples, i.e.:

\begin{multline}\Big<f(\etamacro)
\Big>_{\P^{\rm ABC}( \eta^{{\rm macro}} | \vec d)} = \\ = \int d\etamacro f(\etamacro) \P^{\rm ABC}( \eta^{{\rm macro}} | \vec d) = \\= \frac{1}{K^{\rm acc}}  \sum_{i\in {\rm acc}} f(\etamacro_i)\, .
\end{multline}

For every lens we perform the above iterative scheme a total of three times, inferring the macro-model parameters or the source regularisation level in an alternate fashion.

\subsection{Substructure reconstruction}
\label{sec:substructure_rec}

We now extend our inference to mass substructure within the lensing potential and the relative fraction $\fsub$. 
To this end, we include two additional steps where we first draw values of $\fsub$ from its prior probability distribution $\P(\fsub)$ and then draw explicit substructure realisations $\eta^{\rm sub}$ for each given value of $\fsub$. We draw only one substructure realisation per macro-model. This is justified by the large number of draw samples, which ensures that for each pair ($\etamacro_i$,$\etasub_i$) there exist another one ($\etamacro_j$,$\etasub_j$) with a different substructure realisation but  similar macro-model parameters. For each macro-model and substructure combination we then calculate the Evidence $ \P_i = \P(\vec d| L( \etamacro_i,\etasub_i) )$ and follow the iterative scheme presented above. 

In our approximate posterior we would like to take into account that each substructure realisation could also have been generated from a mass function with a different $\fsub$ than the one it has actually been drawn from. To this end, for each of these realisations we also calculate the probability of having been drawn from a model with $\widetilde \fsub$ and obtain the approximate posterior:
\begin{equation}
 \P^{\rm ABC}(  \fsub | \vec d) = \frac{1}{K^{\rm acc}} \sum_{\rm i \in {\rm acc}} v_i (\fsub ) \,,
\end{equation}
where the interpolating function $v_i$ is defined as:
\begin{equation}
 v_i (\fsub ) = \frac{\P(N^{ \rm sub}_{\rm i}|\fsub) \P(\fsub)}{\int d \widetilde \fsub  ~\P \Big(N^{ \rm sub}_{\rm i}\Big|\widetilde \fsub\Big) \P \Big(\widetilde \fsub\Big)}\,.
\end{equation}
The generated substructure realisations will rarely match the one contained in the data. Rather than that, we expect that realisations drawn from the correct mass function will more often provide higher Evidence than those drawn from a different mass function, as the latter have either too few or too many substructures. If a particular substructure realisation can reproduce the observed lensed images much better than any other, the weights introduced above will  always  ensure that more than a single value of $\fsub$ will contribute to the posterior distribution.
 Using mock data with different level of $\fsub$, we have thoroughly tested our inference approach, and found that we are able to recover the correct amount of mass in substructure with a precision that, as expected, depends on the number of lenses and the data quality.\\
Recently, \cite{despali2018} have shown that low-mass haloes along the line of sight can significantly contribute to the number of detectable objects. For simplicity, here we focus only on the substructure contribution. However, our analysis can be easily extended to include both populations.

\section{Results}
\label{sec:results}

In this section, we present the results of the inference analysis described in Section \ref{sec:reconstruction} when applied to the sample of twenty-four mock lensing data presented in Section \ref{sec:simulations}. In particular, for each lens, we create $N^{\rm step}\times N^{\rm samp}= 20\times4000$ lens configurations and choose $N^{\rm acc}=200$ at each step during the smooth modelling phase and $N^{\rm step}\times N^{\rm samp}= 40\times4000$ lens configurations and accept around half of them for the substructure inference.

\subsection{Smooth modelling}

\begin{figure*}
    \centering
 \includegraphics[width=\hsize]{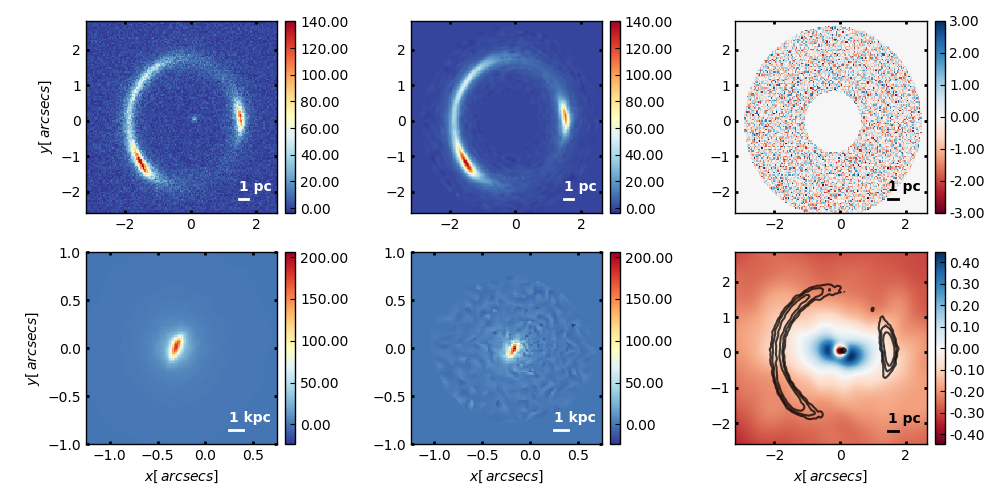}
 \caption{From left to right,  top to bottom:  the mock data $d(\vec x):=I_{\rm gt}(\vec x)$, the reconstructed lensed images $I_{\rm rec}(\vec x)$, the  image-plane residuals $r(\vec x)$, the ground truth source $s_{\rm gt}(\vec x)$, the reconstructed source $s_{\rm rec}(\vec x)$, and the relative difference in convergence $ \Delta \kappa/ \kappa (\vec x) = \kappa_{\rm gt}(\vec x)/ \kappa_{\rm rec} (\vec x) -1$.}
\label{fig:macro_comp1}
\end{figure*}

\begin{figure*}
    \centering
 \includegraphics[width=\hsize]{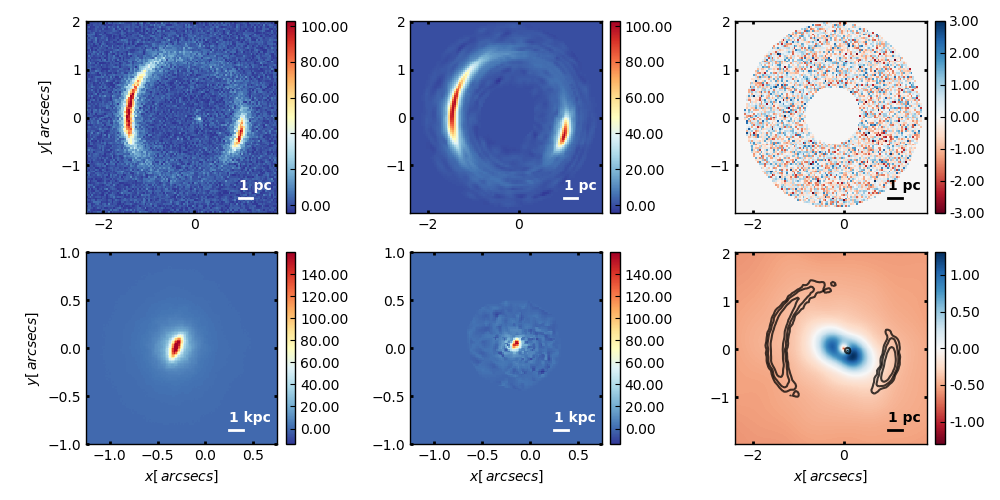}
\caption{Same as Figure \ref{fig:macro_comp1} but for the mock with the highest residuals.}
\label{fig:macro_comp2}
\end{figure*}

\begin{figure}
    \centering
 \includegraphics[width=\hsize]{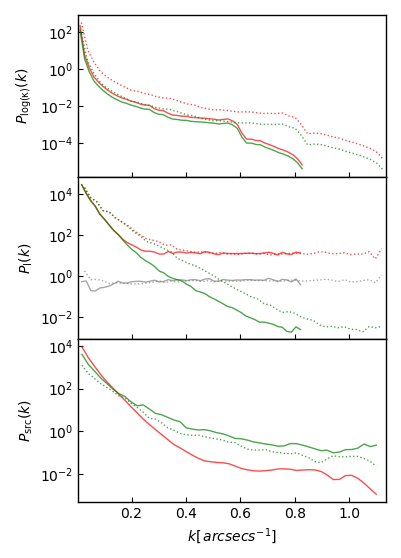}
\caption{The top panel shows the power spectrum of the logarithmic convergence maps $\log(\kappa (\vec x))$ for the groundtruth (red) and the model (green). The middle panel shows the power spectra of the data $d(\vec x)$ (red), the reconstructed lensed images $I(\vec x)$ (green), and the image plane residuals (gray).
The bottom panel shows the power spectra of the groundtruth  source (red) and the reconstructed one (green).
All panels show the spectra for the mock data with the best (solid curve) and worst (dotted curve) reconstruction.}
\label{fig:ps_plot}
\end{figure}

\subsubsection{Lensing mass distribution}

A simple elliptical cored power-law mass distribution (with no substructure) can fit all of the mock observations to the noise level. Figures \ref{fig:macro_comp1} and \ref{fig:macro_comp2} show the best and the worst reconstructions, respectively.
For the worst reconstruction, we find correlated image residuals at the $1.3\sigma$ level, which are related to the high dynamic range of the input lensed images which cannot be fully reproduced by the reconstructed source due to the effect of the regularisation.

Even though we can accurately reproduce the lensed images to the noise level, we infer a lensing mass density distribution with a core that is much smaller than the true value.  
Gravitational lensing is mainly sensitive to the total mass within the Einstein radius and not its specific distribution.
This means that it is insensitive to the presence of a core, when it does not directly affect the lensed images.
As a result, the Illustris-1 galaxies in our sample have a core size of $r_c= $ 0.15-0.5 $\arcsec$, while the reconstructed cores tend to be of the order of $r_c = 5\times10^{-4} \arcsec$. 
As can be seen in the Figures \ref{fig:macro_comp1} and \ref{fig:macro_comp2}, the relative difference in convergence between the simulated lenses and the power-law model stays constant over the extent of the lensed images, and hence can be compensated by a rescaling of the source size. 

In order to further test the influence of the core we have also modelled one of the lens systems where the core directly affects the lensed images (e.g.  middle-bottom panel of \ref{fig:Illustris1_kappa}) and found that in this case we can recover the correct core size within one standard deviation and obtain a source that is of the same size as the original one.

\subsubsection{Source and time delay reconstruction}
In general, we recover sources that tend to be on average 50 per cent more compact than the original ones, as reflected by a lack of power in their power spectra  at large scales of $\mathcal{O}({\rm a~few} \kpc )$. Figure \ref{fig:ps_plot} shows the power spectrum of the logarithmic convergence maps (top), the data (middle) and reconstructed model for the source (bottom); in all panels we can compare the ground truth with the reconstructed model.
This effect on the source size is in agreement with the findings by Ritondale et al. (submitted) who have demonstrated, with the study of real gravitational lens systems, that a failure to fully capture the complexity of the lensing potential leads to a systematic error in the estimated morphological and physical properties of the reconstructed source by up to a factor of 22 per cent.
In particular, we observe a correlation between the logarithm of $P^{\rm \log( \kappa)}_{\rm gt}/ P^{\rm \log( \kappa)}_{\rm lens}(k)~ \forall~ k $ and the logarithm of $P^{\rm s}_{\rm gt}/ P^{\rm s}_{\rm lens}(k)$ for $k = \mathcal{O}( \kpc)$ with a correlation coefficient  $\rho \approx 0.95$ in our sample.
Moreover, we find that the reconstructed sources show increased power at small scales with respect to the ground truth, which indicates a certain level of noise fitting, as also found by \cite{bayer2018}. 
A log-normal source prior may mitigate this issue significantly as it restricts the source brightness values to positive values. This potentially allows for reconstructing sources up to smaller scales and increase the sensitivity to substructures as a result.\\

We further consider the time delays of the reconstructed lens models. 
Following the discussion by \citet{kochanek2019overconstrained} we calculate for each of the lens systems the expected fractional error on $H_0$ using the (mean) convergence at the Einstein radius $\kappa(R_{\rm E})$, i.e. $f_{H_0} = \frac{H_0^{gt}}{H_0^{rec}} -1 = \frac{1- \kappa^{gt}(R_{\rm E})}{1- \kappa^{rec}(R_{\rm E})} -1 =  \frac{ \Delta t_{ij}^{gt} }{ \Delta t_{ij}^{rec}  }-1$, where $\kappa^{gt}_E$ is the ground truth convergence used to generate data, i.e. the Illustris galaxies (see Section \ref{sec:simulations}) or the control sample (see Section \ref{sec:subs_rec_cont}), while $\kappa^{rec}$ is the convergence at the Einstein radius obtained from the lens modelling of these two samples. 
We find a median and $1\sigma$ percentiles of $f_{H_0} = 25_{-19}^{+37}$ per cent and $f_{H_0} = 12_{-3}^{+6}$ per cent, for the Illustris galaxies and the control sample, respectively. In both cases the Hubble constant is, therefore, systematically underestimated. Notice, that our analysis uses as constraints on the mass distribution the extended lensed images only, while cosmographic analyses  \citep[see e.g.][]{suyu2010,chen2019}, also make use of the additional information provided by the kinematics of the lens galaxy and the positions of quadruply-lensed quasars.

\subsection{Substructure reconstruction} \label{sec:subs_rec_cont}

We now allow for the presence of substructure in the lens modelling to infer the amount of lens galaxy mass in substructures, $\fsub$. 

To validate this inference we generate a control sample of twenty-four lens systems with the same observational properties (i.e. signal-to-noise ratio and angular resolution) as our Illustris-1 sample.
To this end, we use as input lensed images those corresponding to the highest evidence macro-model obtained in the previous section by modelling the Illustris-1 galaxies with a smooth power-law. This approach provides us with a sample for which we have a perfect knowledge of the lensing potential.
We model both the mocks generated directly with the Illustris-1 galaxies and the control sample.
As the input data in both samples do not contain any substructure, we expect the control sample to lead to an inferred value of $\fsub$ consistent with zero. Any potential difference in $\fsub$ between the two samples can be then attributed to the additional complexity of the Illustris-1 galaxies.

\begin{figure}
   \centering
  \includegraphics[width=\hsize]{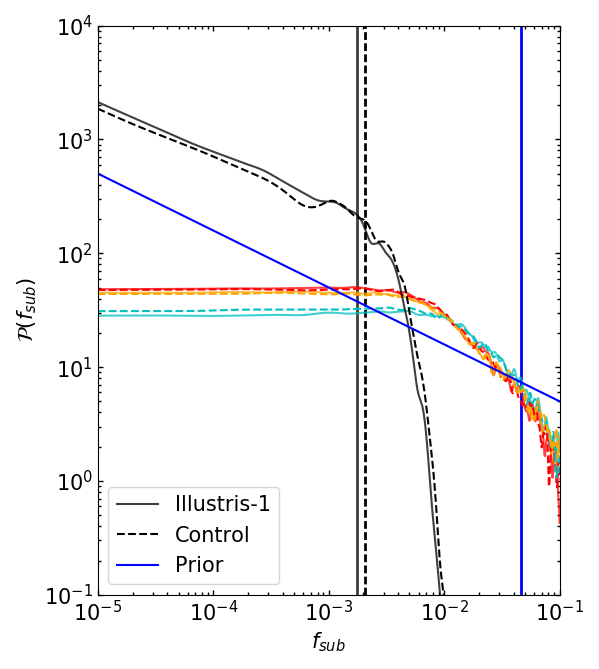}

\caption{The posterior distribution of $\fsub$ for the Illustris-1 sample (continuous curve) and the control sample (dashed curve).   Furthermore, we show the Likelihood for one of the Illustrus-1 lens  galaxies, and show the behaviour for all three projections (red, orange and cyan). . We find that the upper confidence limits of $\fsub$ do not change much between the mock and the control samples (in particular when comparing the reconstructions on the level of individual lens systems). The prior probability distribution of $\fsub$ is shown in blue. The vertical lines indicate the 68 per cent confidence levels (CL) of the prior and posterior distributions for the mock and the control sample respectively.}
\label{fig:fsub_post}
\end{figure}

Figure \ref{fig:fsub_post} shows the joint posterior on $\fsub$ for the Illustris-1 (solid line) and the control sample (dashed line). 
Furthermore, it shows the ABC Likelihood of $\fsub$ for one of the Illustris-1 galaxies in three projections (coloured lines). For small values of $\fsub$ the Likelihood is flat as the number of substructures affecting the lensed surface brightness approaches zero. This flattening may appear before the point where no substructures are placed due to the sensitivity of the data.
Within the region where the Likelihood is flat, the posterior inherits the shape of the prior. For large values of $\fsub$, the Likelihood decreases, reflecting an overabundance of substructures that is not compatible with the data.
 
From the analysis of all gravitational lens systems in each of the two samples, we find a posterior upper limit of $\fsub \leq 1.7\times10^{-3}$ and $\fsub \leq 2\times10^{-3}$ at the $68$ per cent confidence level for the Illustris-1 and the control sample, respectively. The inferred $\fsub$ from the two samples agree with each other within the uncertainty arising from the finite number of samples  of $(\etamacro,\etasub)$ drawn in the analysis. In both cases, the posterior upper limits differ significantly from the prior upper limit of $\fsub \leq 4.7\times10^{-2}$. Both samples are compatible with the Null-Hypothesis that the lensing potential does not contain any substructures, i.e. $\fsub=0$. We conclude, therefore, that baryonic structure, as well as small scale complexity, do not have a significant impact on our capability of correctly inferring the amount of substructure. 
 We find that the sources do not change much with respect to the case without substructure.
In order to test the effect of our choice for the input source galaxy, we repeat the analysis using HST observations of a local galaxy and find no significant change on the inferred upper limit on $\fsub$.

Recently, from the analysis of flux-ratio anomalies in multiply imaged quasars lensed by elliptical galaxies from the Illustris-1 simulation with the same smoothing level, \citet{hsueh2018} have found that baryonic components in the lensing galaxies are responsible for an increase in the probability of flux-ratio anomalies by $8 $ per cent.
Similarly, \citet{gilman2017} have concluded that 10 per cent of the flux-ratio anomalies in elliptical galaxies can be attributed to baryonic structures in their analysis of mock data based on real HST observations. Both results have important implications for the quantification of sub- and line-of-sight structure with gravitationally lensed quasars.
We attribute the difference from our findings to the fact that flux ratios are sensitive to the second derivative of the lensing potential rather than the first, and are therefore more affected by general departures from simple smooth distributions.
As the sensitivity of extended arcs and rings is strongly dependent on the angular resolution of the data, we expect our results to change with increasing data quality.

On the other hand, our results are also in disagreement with \citet{vegetti2014} and Ritondale et al. (submitted), who from the analysis of real lens systems with a similar data quality as ours, have identified the presence of complex mass components which are degenerate with the presence of a large population of substructure and can lead to false detections. This discrepancy seems to indicate that galaxies in the Illustris-1 simulations are smoother than real galaxies and that their level of complexity at scales that we can resolve - i.e. scales  $> \mathcal{O}( kpc) $ - is such that it has a negligible lensing effect which is easily absorbed by changes in the macro-model and the source structure. 
It should also be considered that the simulated galaxies analysed here were chosen to match the global properties of known lenses, but unlike the latter they have not been selected based on their spectral properties. This may potentially lead to a sample of simulated lens galaxies which is more homogeneous and more regular than real ones.
We therefore conclude that higher-resolution simulations together with more realistic selection criteria are key to  properly characterise their central mass distribution in a way that is not affected by the presence of an un-physical core and investigate the level of complexity in more realistic galaxies and its effect on the detection of low-mass haloes.

\section{Summary \& Conclusions}
\label{sec:summary}

Using a novel Bayesian forward modelling technique based on ABC, we have analysed a sample of mock lensing observations generated from a simulated source and lensing potentials taken from the Illustris-1 simulation. Our results have interesting implications for a variety of measurements, from the study of lensed galaxies to constraints on cosmology and dark matter. We summarise them as follows:
\noindent(i) Our capability of reproducing the lensed images down to the noise level without fully correctly modelling the cored central mass density distribution of the lenses indicates some form of the source-position transformation (SPT) \citep{schneider2014}, in line with the previous findings by \citep{unruh2017}. 
As a consequence, our reconstructions lead to a \emph{systematic} fractional error on the Hubble constant of $25_{-19}^{+37}$ per cent (in comparison to a \emph{statical} error of $12_{-3}^{+6}$  per cent when the shape of the lensing potential is perfectly known). This result is in agreement with the latest analysis of \citet{blum2020}, that shows that cored (dark matter) mass density distributions give rise to approximate Mass sheet degeneracies (MSDs), and an error on the inferred Hubble constant. The latest cosmographic analyses \citep[see e.g.][]{wong2019h0licow} have attempted to break these degeneracies by including the information contained in the kinematic properties of the lens galaxies and the positions of the lensed quasar images. However, the validity of this approach has been recently debated by \citet{kochanek2019overconstrained}, who has demonstrated that departures from single power-law mass distributions are responsible for a fractional error on the Hubble constant of 30 per cent.
While the cores in the simulations analysed in this paper are artefacts related to limited resolution, cored mass density distribution in real galaxies may be developed by the effect of baryonic processes \citep[see e.g.][]{Chan_2015} or changes in the dark matter properties \citep{Schive_2014,Spergel_2000}. Moreover, similar additional complexities exist in real galaxies are related, for example, to the presence of faint discs \citep{hsueh2018}, bars, or other (baryonic) structures \citep{gilman2018, xu2013colddarkmatter}. More generally, there exist many plausible deviations from a smooth power-law distribution, such as broken power laws \citep[see e.g.][]{du2019accurate} or multiple component models \citep[see e.g.][]{nightingale2018}, which can produce comparable degeneracies.
Together with the findings of \citet{blum2020}, our results have important implications for the analysis of time delays and a potential solution to the $H_0$-tension \citep{wong2019h0licow}.
\noindent(ii) The MSD we encounter in this work further affects the reconstruction of the sources, for which we underestimate the size by 50 per cent on average. This finding is in agreement with Ritondale et al. (in prep), who showed for a Ly$\alpha$-emitting galaxy of the BELLS GALLERY that a correct modelling of the complex lensing potential 
is essential to reconstruct the properties of the lensed galaxy. In particular, they find that mass components beyond the simple power-law model, in the form of pixelated potential corrections, are required for correct focusing of the source.  
The reconstructed sources also differ from the input source on small scales, reflecting the presence of noise in the data and the choice of a Gaussian regularising
prior. We expect that a log-normal prior or a hyper-prior as in \citep{rizzo2018} may provide more
stable results as well as a better sensitivity to low-mass haloes by reducing the freedom of the source
structure to re-absorb the effect of the latter on the lensed images. 
\noindent(iii) We find no degeneracy between complex structures in the lensing potential and low-mass substructures. After having removed all substructures from the simulations we have inferred a fraction of mass in substructure which is consistent with the absence of any substructure and with what was derived from a control sample for which we have a perfect knowledge of the lensing potential. This is in disagreement with \citet{ritondale2019b}, \citet{xu2015} and \citet{gilman2019}, who have shown both with numerical simulations and observations that un-modelled small scale structures in the lensing mass distribution can lead to significant flux-ratio and surface-brightness anomalies. We believe the origin of this discrepancy to lie partly on the limited resolution of the simulations and partly on the selection criteria adopted in this work, as well a different sensitivity of lensed quasars to changes in the potential.

What is important to notice is that a simple power-law model provides a good fit to the lensing observable. Meaning that the quality of fit is not a good measure to exclude the potential presence of un-modelled mass components. In line with previous works, our results underlines the importance of complex lensing mass distributions that go beyond the standard power-law assumptions as well as the need of extra information \citep[e.g. from stellar kinematics or time delays][]{barnabe2007, schneider2014} to break existing degeneracies. From a numerical perspective, our work shows the importance of higher-resolution and more realistic numerical simulations to further quantify potential systematic effects in strong gravitational lensing reconstructions.

\section*{Acknowledgements}

We thank Matt Auger, Mohammadreza Ayromlou, Francesca Rizzo, Elisa Ritondale, Devon Powell, Jens Jasche, Sampath Mukherjee, Matteo Frigo, Simon Birrer and Daniel Gilman for insightful discussions. Simona Vegetti has received funding from the European Research Council (ERC) under the European Union's Horizon 2020 research and innovation programme (grant agreement No 758853).
We also want to thank the Illustris-Collaboration for making their data publically available.



\bibliographystyle{mnras}
\bibliography{biblio.bib}

\begin{thebibliography}{}
\makeatletter
\relax
\def\mn@urlcharsother{\let\do\@makeother \do\$\do\&\do\#\do\^\do\_\do\%\do\~}
\def\mn@doi{\begingroup\mn@urlcharsother \@ifnextchar [ {\mn@doi@}
  {\mn@doi@[]}}
\def\mn@doi@[#1]#2{\def\@tempa{#1}\ifx\@tempa\@empty \href
  {http://dx.doi.org/#2} {doi:#2}\else \href {http://dx.doi.org/#2} {#1}\fi
  \endgroup}
\def\mn@eprint#1#2{\mn@eprint@#1:#2::\@nil}
\def\mn@eprint@arXiv#1{\href {http://arxiv.org/abs/#1} {{\tt arXiv:#1}}}
\def\mn@eprint@dblp#1{\href {http://dblp.uni-trier.de/rec/bibtex/#1.xml}
  {dblp:#1}}
\def\mn@eprint@#1:#2:#3:#4\@nil{\def\@tempa {#1}\def\@tempb {#2}\def\@tempc
  {#3}\ifx \@tempc \@empty \let \@tempc \@tempb \let \@tempb \@tempa \fi \ifx
  \@tempb \@empty \def\@tempb {arXiv}\fi \@ifundefined
  {mn@eprint@\@tempb}{\@tempb:\@tempc}{\expandafter \expandafter \csname
  mn@eprint@\@tempb\endcsname \expandafter{\@tempc}}}

\bibitem[\protect\citeauthoryear{{Akeret}, {Refregier}, {Amara}, {Seehars}  \&
  {Hasner}}{{Akeret} et~al.}{2015}]{akaret2015}
{Akeret} J.,  {Refregier} A.,  {Amara} A.,  {Seehars} S.,   {Hasner} C.,  2015,
  \mn@doi [Journal of Cosmology and Astro-Particle Physics]
  {10.1088/1475-7516/2015/08/043}, \href
  {https://ui.adsabs.harvard.edu/\#abs/2015JCAP...08..043A} {2015, 043}

\bibitem[\protect\citeauthoryear{{Barkana}}{{Barkana}}{1998}]{barkana1998}
{Barkana} R.,  1998, \mn@doi [\apj] {10.1086/305950}, \href
  {https://ui.adsabs.harvard.edu/\#abs/1998ApJ...502..531B} {502, 531}

\bibitem[\protect\citeauthoryear{{Barnab{\`e}} \& {Koopmans}}{{Barnab{\`e}} \&
  {Koopmans}}{2007}]{barnabe2007}
{Barnab{\`e}} M.,  {Koopmans} L. V.~E.,  2007, \mn@doi [\apj] {10.1086/520495},
  \href {https://ui.adsabs.harvard.edu/abs/2007ApJ...666..726B} {666, 726}

\bibitem[\protect\citeauthoryear{{Bayer}, {Chatterjee}, {Koopmans}, {Vegetti},
  {McKean}, {Treu}  \& {Fassnacht}}{{Bayer} et~al.}{2018}]{bayer2018}
{Bayer} D.,  {Chatterjee} S.,  {Koopmans} L.~V.~E.,  {Vegetti} S.,  {McKean}
  J.~P.,  {Treu} T.,   {Fassnacht} C.~D.,  2018, arXiv e-prints, \href
  {https://ui.adsabs.harvard.edu/abs/2018arXiv180305952B} {p. arXiv:1803.05952}

\bibitem[\protect\citeauthoryear{{Birrer}, {Amara}  \& {Refregier}}{{Birrer}
  et~al.}{2017}]{birrer2017}
{Birrer} S.,  {Amara} A.,   {Refregier} A.,  2017, \mn@doi [Journal of
  Cosmology and Astro-Particle Physics] {10.1088/1475-7516/2017/05/037}, \href
  {https://ui.adsabs.harvard.edu/\#abs/2017JCAP...05..037B} {2017, 037}

\bibitem[\protect\citeauthoryear{Blum, Castorina  \& Simonovi?}{Blum
  et~al.}{2020}]{blum2020}
Blum K.,  Castorina E.,   Simonovi? M.,  2020, Could quasar lensing time delays
  hint to cored dark matter halos, instead of $H_0$ tension? (\mn@eprint
  {arXiv} {2001.07182})

\bibitem[\protect\citeauthoryear{{Bolton}, {Burles}, {Koopmans}, {Treu}  \&
  {Moustakas}}{{Bolton} et~al.}{2006}]{Bolton06}
{Bolton} A.~S.,  {Burles} S.,  {Koopmans} L.~V.~E.,  {Treu} T.,   {Moustakas}
  L.~A.,  2006, \mn@doi [\apj] {10.1086/498884}, \href
  {http://adsabs.harvard.edu/abs/2006ApJ...638..703B} {638, 703}

\bibitem[\protect\citeauthoryear{Brewer, Huijser  \& Lewis}{Brewer
  et~al.}{2015}]{brewer2015}
Brewer B.~J.,  Huijser D.,   Lewis G.~F.,  2015, \mn@doi [Monthly Notices of
  the Royal Astronomical Society] {10.1093/mnras/stv2370}, 455, 1819

\bibitem[\protect\citeauthoryear{{Bullock}, {Kolatt}, {Sigad}, {Somerville},
  {Kravtsov}, {Klypin}, {Primack}  \& {Dekel}}{{Bullock}
  et~al.}{2001}]{bullock2001}
{Bullock} J.~S.,  {Kolatt} T.~S.,  {Sigad} Y.,  {Somerville} R.~S.,  {Kravtsov}
  A.~V.,  {Klypin} A.~A.,  {Primack} J.~R.,   {Dekel} A.,  2001, \mn@doi
  [\mnras] {10.1046/j.1365-8711.2001.04068.x}, \href
  {http://adsabs.harvard.edu/abs/2001MNRAS.321..559B} {321, 559}

\bibitem[\protect\citeauthoryear{Chan, Kere¨, Oñorbe, Hopkins, Muratov,
  Faucher-Giguère  \& Quataert}{Chan et~al.}{2015}]{Chan_2015}
Chan T.~K.,  Kere¨ D.,  Oñorbe J.,  Hopkins P.~F.,  Muratov A.~L.,
  Faucher-Giguère C.-A.,   Quataert E.,  2015, \mn@doi [Monthly Notices of the
  Royal Astronomical Society] {10.1093/mnras/stv2165}, 454, 2981?3001

\bibitem[\protect\citeauthoryear{Chen et~al.,}{Chen et~al.}{2019}]{chen2019}
Chen G. C.-F.,  et~al., 2019, \mn@doi [Monthly Notices of the Royal
  Astronomical Society] {10.1093/mnras/stz2547}

\bibitem[\protect\citeauthoryear{{Cox}}{{Cox}}{1946}]{cox1946}
{Cox} R.~T.,  1946, \mn@doi [American Journal of Physics] {10.1119/1.1990764},
  \href {https://ui.adsabs.harvard.edu/abs/1946AmJPh..14....1C} {14, 1}

\bibitem[\protect\citeauthoryear{Daylan, Cyr-Racine, Rivero, Dvorkin  \&
  Finkbeiner}{Daylan et~al.}{2018}]{daylan2018}
Daylan T.,  Cyr-Racine F.-Y.,  Rivero A.~D.,  Dvorkin C.,   Finkbeiner D.~P.,
  2018, \mn@doi [The Astrophysical Journal] {10.3847/1538-4357/aaaa1e}, 854,
  141

\bibitem[\protect\citeauthoryear{{Despali} \& {Vegetti}}{{Despali} \&
  {Vegetti}}{2017}]{despali2017}
{Despali} G.,  {Vegetti} S.,  2017, \mn@doi [\mnras] {10.1093/mnras/stx966},
  \href {https://ui.adsabs.harvard.edu/\#abs/2017MNRAS.469.1997D} {469, 1997}

\bibitem[\protect\citeauthoryear{{Despali}, {Vegetti}, {White}, {Giocoli}  \&
  {van den Bosch}}{{Despali} et~al.}{2018}]{despali2018}
{Despali} G.,  {Vegetti} S.,  {White} S. D.~M.,  {Giocoli} C.,   {van den
  Bosch} F.~C.,  2018, \mn@doi [\mnras] {10.1093/mnras/sty159}, \href
  {https://ui.adsabs.harvard.edu/\#abs/2018MNRAS.475.5424D} {475, 5424}

\bibitem[\protect\citeauthoryear{Du, Zhao, Fan, Shu, Li  \& Mao}{Du
  et~al.}{2019}]{du2019accurate}
Du W.,  Zhao G.-B.,  Fan Z.,  Shu Y.,  Li R.,   Mao S.,  2019, An accurate
  analytic model for the lensing mass of galaxies (\mn@eprint {arXiv}
  {1911.11761})

\bibitem[\protect\citeauthoryear{{Duffy}, {Schaye}, {Kay}  \& {Dalla
  Vecchia}}{{Duffy} et~al.}{2008}]{duffy2008}
{Duffy} A.~R.,  {Schaye} J.,  {Kay} S.~T.,   {Dalla Vecchia} C.,  2008, \mn@doi
  [\mnras] {10.1111/j.1745-3933.2008.00537.x}, \href
  {http://adsabs.harvard.edu/abs/2008MNRAS.390L..64D} {390, L64}

\bibitem[\protect\citeauthoryear{{Falco}, {Gorenstein}  \& {Shapiro}}{{Falco}
  et~al.}{1985}]{falco1985}
{Falco} E.~E.,  {Gorenstein} M.~V.,   {Shapiro} I.~I.,  1985, \mn@doi [\apjl]
  {10.1086/184422}, \href
  {https://ui.adsabs.harvard.edu/abs/1985ApJ...289L...1F} {289, L1}

\bibitem[\protect\citeauthoryear{{Gao}, {Frenk}, {Boylan-Kolchin}, {Jenkins},
  {Springel}  \& {White}}{{Gao} et~al.}{2011}]{gao2011}
{Gao} L.,  {Frenk} C.~S.,  {Boylan-Kolchin} M.,  {Jenkins} A.,  {Springel} V.,
   {White} S.~D.~M.,  2011, \mn@doi [\mnras]
  {10.1111/j.1365-2966.2010.17601.x}, \href
  {http://adsabs.harvard.edu/abs/2011MNRAS.410.2309G} {410, 2309}

\bibitem[\protect\citeauthoryear{{Genel}, {Fall}, {Hernquist}, {Vogelsberger},
  {Snyder}, {Rodriguez-Gomez}, {Sijacki}  \& {Springel}}{{Genel}
  et~al.}{2015}]{genel2015}
{Genel} S.,  {Fall} S.~M.,  {Hernquist} L.,  {Vogelsberger} M.,  {Snyder}
  G.~F.,  {Rodriguez-Gomez} V.,  {Sijacki} D.,   {Springel} V.,  2015, \mn@doi
  [\apjl] {10.1088/2041-8205/804/2/L40}, \href
  {http://adsabs.harvard.edu/abs/2015ApJ...804L..40G} {804, L40}

\bibitem[\protect\citeauthoryear{{Gilman}, {Agnello}, {Treu}, {Keeton}  \&
  {Nierenberg}}{{Gilman} et~al.}{2017}]{gilman2017}
{Gilman} D.,  {Agnello} A.,  {Treu} T.,  {Keeton} C.~R.,   {Nierenberg} A.~M.,
  2017, \mn@doi [\mnras] {10.1093/mnras/stx158}, \href
  {https://ui.adsabs.harvard.edu/abs/2017MNRAS.467.3970G} {467, 3970}

\bibitem[\protect\citeauthoryear{{Gilman}, {Birrer}, {Treu}, {Keeton}  \&
  {Nierenberg}}{{Gilman} et~al.}{2018}]{gilman2018}
{Gilman} D.,  {Birrer} S.,  {Treu} T.,  {Keeton} C.~R.,   {Nierenberg} A.,
  2018, \mn@doi [\mnras] {10.1093/mnras/sty2261}, \href
  {https://ui.adsabs.harvard.edu/abs/2018MNRAS.481..819G} {481, 819}

\bibitem[\protect\citeauthoryear{{Gilman}, {Birrer}, {Treu}, {Nierenberg}  \&
  {Benson}}{{Gilman} et~al.}{2019}]{gilman2019}
{Gilman} D.,  {Birrer} S.,  {Treu} T.,  {Nierenberg} A.,   {Benson} A.,  2019,
  arXiv e-prints, \href
  {https://ui.adsabs.harvard.edu/\#abs/2019arXiv190111031G} {p.
  arXiv:1901.11031}

\bibitem[\protect\citeauthoryear{{Hezaveh} et~al.,}{{Hezaveh}
  et~al.}{2016}]{hezaveh2016}
{Hezaveh} Y.~D.,  et~al., 2016, \mn@doi [\apj] {10.3847/0004-637X/823/1/37},
  \href {http://adsabs.harvard.edu/abs/2016ApJ...823...37H} {823, 37}

\bibitem[\protect\citeauthoryear{{Hinshaw} et~al.,}{{Hinshaw}
  et~al.}{2013}]{hinshaw2013}
{Hinshaw} G.,  et~al., 2013, \mn@doi [\apjs] {10.1088/0067-0049/208/2/19},
  \href {http://adsabs.harvard.edu/abs/2013ApJS..208...19H} {208, 19}

\bibitem[\protect\citeauthoryear{{Hsueh}, {Fassnacht}, {Vegetti}, {McKean},
  {Spingola}, {Auger}, {Koopmans}  \& {Lagattuta}}{{Hsueh}
  et~al.}{2016}]{hsueh2016}
{Hsueh} J.~W.,  {Fassnacht} C.~D.,  {Vegetti} S.,  {McKean} J.~P.,  {Spingola}
  C.,  {Auger} M.~W.,  {Koopmans} L.~V.~E.,   {Lagattuta} D.~J.,  2016, \mn@doi
  [\mnras] {10.1093/mnrasl/slw146}, \href
  {https://ui.adsabs.harvard.edu/abs/2016MNRAS.463L..51H} {463, L51}

\bibitem[\protect\citeauthoryear{{Hsueh} et~al.,}{{Hsueh}
  et~al.}{2017}]{hsueh2017}
{Hsueh} J.~W.,  et~al., 2017, \mn@doi [\mnras] {10.1093/mnras/stx1082}, \href
  {https://ui.adsabs.harvard.edu/\#abs/2017MNRAS.469.3713H} {469, 3713}

\bibitem[\protect\citeauthoryear{{Hsueh}, {Despali}, {Vegetti}, {Xu},
  {Fassnacht}  \& {Metcalf}}{{Hsueh} et~al.}{2018}]{hsueh2018}
{Hsueh} J.-W.,  {Despali} G.,  {Vegetti} S.,  {Xu} D.~a.,  {Fassnacht} C.~D.,
  {Metcalf} R.~B.,  2018, \mn@doi [\mnras] {10.1093/mnras/stx3320}, \href
  {https://ui.adsabs.harvard.edu/abs/2018MNRAS.475.2438H} {475, 2438}

\bibitem[\protect\citeauthoryear{{Hsueh}, {Enzi}, {Vegetti}, {Auger},
  {Fassnacht}, {Despali}, {Koopmans}  \& {McKean}}{{Hsueh}
  et~al.}{2019}]{hsueh2019}
{Hsueh} J.-W.,  {Enzi} W.,  {Vegetti} S.,  {Auger} M.,  {Fassnacht} C.~D.,
  {Despali} G.,  {Koopmans} L. V.~E.,   {McKean} J.~P.,  2019, arXiv e-prints,
  \href {https://ui.adsabs.harvard.edu/abs/2019arXiv190504182H} {p.
  arXiv:1905.04182}

\bibitem[\protect\citeauthoryear{{Jeffreys}}{{Jeffreys}}{1946}]{jeffrey1946}
{Jeffreys} H.,  1946, \mn@doi [Proceedings of the Royal Society of London
  Series A] {10.1098/rspa.1946.0056}, \href
  {https://ui.adsabs.harvard.edu/abs/1946RSPSA.186..453J} {186, 453}

\bibitem[\protect\citeauthoryear{Kochanek}{Kochanek}{2019}]{kochanek2019overconstrained}
Kochanek C.~S.,  2019, Over-constrained Gravitational Lens Models and the
  Hubble Constant (\mn@eprint {arXiv} {1911.05083})

\bibitem[\protect\citeauthoryear{{Koopmans}}{{Koopmans}}{2005}]{koopmans2005}
{Koopmans} L.~V.~E.,  2005, \mn@doi [\mnras]
  {10.1111/j.1365-2966.2005.09523.x}, \href
  {http://adsabs.harvard.edu/abs/2005MNRAS.363.1136K} {363, 1136}

\bibitem[\protect\citeauthoryear{{Kormann}, {Schneider}  \&
  {Bartelmann}}{{Kormann} et~al.}{1994}]{kormann1994}
{Kormann} R.,  {Schneider} P.,   {Bartelmann} M.,  1994, \aap, \href
  {https://ui.adsabs.harvard.edu/\#abs/1994A&A...284..285K} {284, 285}

\bibitem[\protect\citeauthoryear{{Mao} \& {Schneider}}{{Mao} \&
  {Schneider}}{1998}]{mao1998}
{Mao} S.,  {Schneider} P.,  1998, \mn@doi [\mnras]
  {10.1046/j.1365-8711.1998.01319.x}, \href
  {http://adsabs.harvard.edu/abs/1998MNRAS.295..587M} {295, 587}

\bibitem[\protect\citeauthoryear{{Metcalf} \& {Petkova}}{{Metcalf} \&
  {Petkova}}{2014}]{metcalf2014}
{Metcalf} R.~B.,  {Petkova} M.,  2014, \mn@doi [\mnras]
  {10.1093/mnras/stu1859}, \href
  {https://ui.adsabs.harvard.edu/\#abs/2014MNRAS.445.1942M} {445, 1942}

\bibitem[\protect\citeauthoryear{{M{\"o}ller}, {Hewett}  \&
  {Blain}}{{M{\"o}ller} et~al.}{2003}]{moeller2003}
{M{\"o}ller} O.,  {Hewett} P.,   {Blain} A.~W.,  2003, \mn@doi [\mnras]
  {10.1046/j.1365-8711.2003.06758.x}, \href
  {https://ui.adsabs.harvard.edu/abs/2003MNRAS.345....1M} {345, 1}

\bibitem[\protect\citeauthoryear{{Mukherjee} et~al.,}{{Mukherjee}
  et~al.}{2018}]{mukherjee2018}
{Mukherjee} S.,  et~al., 2018, \mn@doi [\mnras] {10.1093/mnras/sty1741}, \href
  {https://ui.adsabs.harvard.edu/abs/2018MNRAS.479.4108M} {479, 4108}

\bibitem[\protect\citeauthoryear{{Navarro}, {Frenk}  \& {White}}{{Navarro}
  et~al.}{1997}]{navarro1996}
{Navarro} J.~F.,  {Frenk} C.~S.,   {White} S. D.~M.,  1997, \mn@doi [\apj]
  {10.1086/304888}, \href
  {https://ui.adsabs.harvard.edu/abs/1997ApJ...490..493N} {490, 493}

\bibitem[\protect\citeauthoryear{{Nierenberg}, {Treu}, {Wright}, {Fassnacht}
  \& {Auger}}{{Nierenberg} et~al.}{2014}]{nierenberg2014}
{Nierenberg} A.~M.,  {Treu} T.,  {Wright} S.~A.,  {Fassnacht} C.~D.,   {Auger}
  M.~W.,  2014, \mn@doi [\mnras] {10.1093/mnras/stu862}, \href
  {https://ui.adsabs.harvard.edu/abs/2014MNRAS.442.2434N} {442, 2434}

\bibitem[\protect\citeauthoryear{Nightingale, Dye  \& Massey}{Nightingale
  et~al.}{2018}]{nightingale2018}
Nightingale J.~W.,  Dye S.,   Massey R.~J.,  2018, \mn@doi [Monthly Notices of
  the Royal Astronomical Society] {10.1093/mnras/sty1264}, 478, 4738?4784

\bibitem[\protect\citeauthoryear{{Petkova}, {Metcalf}  \& {Giocoli}}{{Petkova}
  et~al.}{2014}]{petkova2014}
{Petkova} M.,  {Metcalf} R.~B.,   {Giocoli} C.,  2014, \mn@doi [\mnras]
  {10.1093/mnras/stu1860}, \href
  {https://ui.adsabs.harvard.edu/\#abs/2014MNRAS.445.1954P} {445, 1954}

\bibitem[\protect\citeauthoryear{{Ritondale}, {Vegetti}, {Despali}, {Auger},
  {Koopmans}  \& {McKean}}{{Ritondale} et~al.}{2019}]{ritondale2019b}
{Ritondale} E.,  {Vegetti} S.,  {Despali} G.,  {Auger} M.~W.,  {Koopmans}
  L.~V.~E.,   {McKean} J.~P.,  2019, \mn@doi [\mnras] {10.1093/mnras/stz464},
  \href {https://ui.adsabs.harvard.edu/abs/2019MNRAS.485.2179R} {485, 2179}

\bibitem[\protect\citeauthoryear{{Rizzo}, {Vegetti}, {Fraternali}  \& {Di
  Teodoro}}{{Rizzo} et~al.}{2018}]{rizzo2018}
{Rizzo} F.,  {Vegetti} S.,  {Fraternali} F.,   {Di Teodoro} E.,  2018, \mn@doi
  [\mnras] {10.1093/mnras/sty2594}, \href
  {https://ui.adsabs.harvard.edu/abs/2018MNRAS.481.5606R} {481, 5606}

\bibitem[\protect\citeauthoryear{{Rybak}, {Vegetti}, {McKean}, {Andreani}  \&
  {White}}{{Rybak} et~al.}{2015}]{rybak2015}
{Rybak} M.,  {Vegetti} S.,  {McKean} J.~P.,  {Andreani} P.,   {White} S.~D.~M.,
   2015, \mn@doi [\mnras] {10.1093/mnrasl/slv092}, \href
  {https://ui.adsabs.harvard.edu/abs/2015MNRAS.453L..26R} {453, L26}

\bibitem[\protect\citeauthoryear{{Schaye} et~al.,}{{Schaye}
  et~al.}{2015}]{schaye2015}
{Schaye} J.,  et~al., 2015, \mn@doi [\mnras] {10.1093/mnras/stu2058}, \href
  {https://ui.adsabs.harvard.edu/abs/2015MNRAS.446..521S} {446, 521}

\bibitem[\protect\citeauthoryear{Schive, Chiueh  \& Broadhurst}{Schive
  et~al.}{2014}]{Schive_2014}
Schive H.-Y.,  Chiueh T.,   Broadhurst T.,  2014, \mn@doi [Nature Physics]
  {10.1038/nphys2996}, 10, 496?499

\bibitem[\protect\citeauthoryear{{Schneider} \& {Sluse}}{{Schneider} \&
  {Sluse}}{2014}]{schneider2014}
{Schneider} P.,  {Sluse} D.,  2014, \mn@doi [\aap]
  {10.1051/0004-6361/201322106}, \href
  {https://ui.adsabs.harvard.edu/abs/2014A&A...564A.103S} {564, A103}

\bibitem[\protect\citeauthoryear{Shirazi, Vegetti, Nesvadba, Allam, Brinchmann
  \& Tucker}{Shirazi et~al.}{2014}]{shirazi2014}
Shirazi M.,  Vegetti S.,  Nesvadba N.,  Allam S.,  Brinchmann J.,   Tucker D.,
  2014, \mn@doi [Monthly Notices of the Royal Astronomical Society]
  {10.1093/mnras/stu316}, 440, 2201

\bibitem[\protect\citeauthoryear{Spergel \& Steinhardt}{Spergel \&
  Steinhardt}{2000}]{Spergel_2000}
Spergel D.~N.,  Steinhardt P.~J.,  2000, \mn@doi [Physical Review Letters]
  {10.1103/physrevlett.84.3760}, 84, 3760?3763

\bibitem[\protect\citeauthoryear{Spingola, McKean, Massari  \&
  Koopmans}{Spingola et~al.}{2019}]{spingola2019}
Spingola C.,  McKean J.~P.,  Massari D.,   Koopmans L. V.~E.,  2019, \mn@doi
  [Astronomy & Astrophysics] {10.1051/0004-6361/201935427}, 630, A108

\bibitem[\protect\citeauthoryear{{Springel}}{{Springel}}{2010}]{springel10}
{Springel} V.,  2010, \mn@doi [\mnras] {10.1111/j.1365-2966.2009.15715.x},
  \href {http://adsabs.harvard.edu/abs/2010MNRAS.401..791S} {401, 791}

\bibitem[\protect\citeauthoryear{{Springel}, {White}, {Tormen}  \&
  {Kauffmann}}{{Springel} et~al.}{2001}]{springel2001}
{Springel} V.,  {White} S. D.~M.,  {Tormen} G.,   {Kauffmann} G.,  2001,
  \mn@doi [\mnras] {10.1046/j.1365-8711.2001.04912.x}, \href
  {https://ui.adsabs.harvard.edu/\#abs/2001MNRAS.328..726S} {328, 726}

\bibitem[\protect\citeauthoryear{{Springel} et~al.,}{{Springel}
  et~al.}{2008}]{Springel2008}
{Springel} V.,  et~al., 2008, \mn@doi [\mnras]
  {10.1111/j.1365-2966.2008.14066.x}, \href
  {http://adsabs.harvard.edu/abs/2008MNRAS.391.1685S} {391, 1685}

\bibitem[\protect\citeauthoryear{{Suyu}, {Marshall}, {Hobson}  \&
  {Blandford}}{{Suyu} et~al.}{2006}]{suyu2006}
{Suyu} S.~H.,  {Marshall} P.~J.,  {Hobson} M.~P.,   {Blandford} R.~D.,  2006,
  \mn@doi [\mnras] {10.1111/j.1365-2966.2006.10733.x}, \href
  {https://ui.adsabs.harvard.edu/\#abs/2006MNRAS.371..983S} {371, 983}

\bibitem[\protect\citeauthoryear{Suyu, Marshall, Auger, Hilbert, Blandford,
  Koopmans, Fassnacht  \& Treu}{Suyu et~al.}{2010}]{suyu2010}
Suyu S.~H.,  Marshall P.~J.,  Auger M.~W.,  Hilbert S.,  Blandford R.~D.,
  Koopmans L. V.~E.,  Fassnacht C.~D.,   Treu T.,  2010, \mn@doi [The
  Astrophysical Journal] {10.1088/0004-637x/711/1/201}, 711, 201

\bibitem[\protect\citeauthoryear{{Teklu}, {Remus}, {Dolag}, {Beck}, {Burkert},
  {Schmidt}, {Schulze}  \& {Steinborn}}{{Teklu} et~al.}{2015}]{teklu15}
{Teklu} A.~F.,  {Remus} R.-S.,  {Dolag} K.,  {Beck} A.~M.,  {Burkert} A.,
  {Schmidt} A.~S.,  {Schulze} F.,   {Steinborn} L.~K.,  2015, \mn@doi [\apj]
  {10.1088/0004-637X/812/1/29}, \href
  {http://adsabs.harvard.edu/abs/2015ApJ...812...29T} {812, 29}

\bibitem[\protect\citeauthoryear{Treu}{Treu}{2010}]{treu2010}
Treu T.,  2010, \mn@doi [Annual Review of Astronomy and Astrophysics]
  {10.1146/annurev-astro-081309-130924}, 48, 87

\bibitem[\protect\citeauthoryear{{Unruh}, {Schneider}  \& {Sluse}}{{Unruh}
  et~al.}{2017}]{unruh2017}
{Unruh} S.,  {Schneider} P.,   {Sluse} D.,  2017, \mn@doi [\aap]
  {10.1051/0004-6361/201629048}, \href
  {https://ui.adsabs.harvard.edu/abs/2017A&A...601A..77U} {601, A77}

\bibitem[\protect\citeauthoryear{{Vegetti} \& {Koopmans}}{{Vegetti} \&
  {Koopmans}}{2009a}]{vegetti2009_2}
{Vegetti} S.,  {Koopmans} L.~V.~E.,  2009a, \mn@doi [\mnras]
  {10.1111/j.1365-2966.2008.14005.x}, \href
  {http://adsabs.harvard.edu/abs/2009MNRAS.392..945V} {392, 945}

\bibitem[\protect\citeauthoryear{{Vegetti} \& {Koopmans}}{{Vegetti} \&
  {Koopmans}}{2009b}]{vegetti2009}
{Vegetti} S.,  {Koopmans} L.~V.~E.,  2009b, \mn@doi [\mnras]
  {10.1111/j.1365-2966.2009.15559.x}, \href
  {https://ui.adsabs.harvard.edu/\#abs/2009MNRAS.400.1583V} {400, 1583}

\bibitem[\protect\citeauthoryear{{Vegetti}, {Koopmans}, {Bolton}, {Treu}  \&
  {Gavazzi}}{{Vegetti} et~al.}{2010}]{vegetti2010b}
{Vegetti} S.,  {Koopmans} L.~V.~E.,  {Bolton} A.,  {Treu} T.,   {Gavazzi} R.,
  2010, \mn@doi [\mnras] {10.1111/j.1365-2966.2010.16865.x}, \href
  {https://ui.adsabs.harvard.edu/abs/2010MNRAS.408.1969V} {408, 1969}

\bibitem[\protect\citeauthoryear{{Vegetti}, {Lagattuta}, {McKean}, {Auger},
  {Fassnacht}  \& {Koopmans}}{{Vegetti} et~al.}{2012}]{vegetti2012}
{Vegetti} S.,  {Lagattuta} D.~J.,  {McKean} J.~P.,  {Auger} M.~W.,  {Fassnacht}
  C.~D.,   {Koopmans} L.~V.~E.,  2012, \mn@doi [\nat] {10.1038/nature10669},
  \href {https://ui.adsabs.harvard.edu/abs/2012Natur.481..341V} {481, 341}

\bibitem[\protect\citeauthoryear{{Vegetti}, {Koopmans}, {Auger}, {Treu}  \&
  {Bolton}}{{Vegetti} et~al.}{2014}]{vegetti2014}
{Vegetti} S.,  {Koopmans} L.~V.~E.,  {Auger} M.~W.,  {Treu} T.,   {Bolton}
  A.~S.,  2014, \mn@doi [\mnras] {10.1093/mnras/stu943}, \href
  {https://ui.adsabs.harvard.edu/\#abs/2014MNRAS.442.2017V} {442, 2017}

\bibitem[\protect\citeauthoryear{{Vogelsberger} et~al.,}{{Vogelsberger}
  et~al.}{2014}]{vogelsberger2014}
{Vogelsberger} M.,  et~al., 2014, \mn@doi [\mnras] {10.1093/mnras/stu1536},
  \href {https://ui.adsabs.harvard.edu/\#abs/2014MNRAS.444.1518V} {444, 1518}

\bibitem[\protect\citeauthoryear{Wong et~al.,}{Wong
  et~al.}{2019b}]{wong2019h0licow}
Wong K.~C.,  et~al., 2019b, H0LiCOW XIII. A 2.4
  lensed quasars: $5.3?$ tension between early and late-Universe probes
  (\mn@eprint {arXiv} {1907.04869})

\bibitem[\protect\citeauthoryear{Wong et~al.,}{Wong et~al.}{2019a}]{wong2019}
Wong K.~C.,  et~al., 2019a, H0LiCOW XIII. A 2.4
  lensed quasars: $5.3σ$ tension between early and late-Universe probes
  (\mn@eprint {arXiv} {1907.04869})

\bibitem[\protect\citeauthoryear{Xu, Sluse, Gao, Wang, Frenk, Mao  \&
  Schneider}{Xu et~al.}{2013}]{xu2013colddarkmatter}
Xu D.~D.,  Sluse D.,  Gao L.,  Wang J.,  Frenk C.,  Mao S.,   Schneider P.,
  2013, How well can cold-dark-matter substructures account for the observed
  lensing flux-ratio anomalies? (\mn@eprint {arXiv} {1307.4220})

\bibitem[\protect\citeauthoryear{{Xu}, {Sluse}, {Gao}, {Wang}, {Frenk}, {Mao},
  {Schneider}  \& {Springel}}{{Xu} et~al.}{2015}]{xu2015}
{Xu} D.,  {Sluse} D.,  {Gao} L.,  {Wang} J.,  {Frenk} C.,  {Mao} S.,
  {Schneider} P.,   {Springel} V.,  2015, \mn@doi [\mnras]
  {10.1093/mnras/stu2673}, \href
  {http://adsabs.harvard.edu/abs/2015MNRAS.447.3189X} {447, 3189}

\bibitem[\protect\citeauthoryear{{Xu}, {Springel}, {Sluse}, {Schneider},
  {Sonnenfeld}, {Nelson}, {Vogelsberger}  \& {Hernquist}}{{Xu}
  et~al.}{2017}]{xu2017}
{Xu} D.,  {Springel} V.,  {Sluse} D.,  {Schneider} P.,  {Sonnenfeld} A.,
  {Nelson} D.,  {Vogelsberger} M.,   {Hernquist} L.,  2017, \mn@doi [Monthly
  Notices of the Royal Astronomical Society] {10.1093/mnras/stx899}, \href
  {https://ui.adsabs.harvard.edu/abs/2017MNRAS.469.1824X} {469, 1824}

\makeatother
\end{thebibliography}



\appendix

\section{The Einstein radius}
\label{sec:einstein_radius}

The Einstein radius $ \RE$ is defined as the radius of a circle within which the average mass density is given by the critical density:

\begin{equation}
    M(<  \RE)= A(<\RE)\times \Sigma_{\rm crit} = \pi \RE^2 \times \Sigma_{\rm crit} \,.
\end{equation}

In this case, it is straightforward to calculate the mass within the Einstein radius by integrating the parameterised convergence within the circle and multiplying it by $\Sigma_{\rm crit}$. Solving the resulting equation for $\RE$ yields a definition of the Einstein radius in the chosen parameterisation of the convergence.
In order to obtain the elliptical Einstein radius we consider the rescaled radius $ \RE\rightarrow  \widetilde \RE = \RE / \sqrt{q} $, so that:

\begin{equation}
    M \left(<  \widetilde \RE \right) =  \pi \widetilde \RE^2 \times \Sigma_{\rm crit} \,,
\end{equation}
which is consistent with the previous definition for $q=1$.
For the parameterized version of the macro-model used in this work, we determine the mass within $\RE= \widetilde \RE /\sqrt{q}$ as: 

\begin{multline}
  M(< \RE/ \sqrt{q} )=  \\ = \Sigma_{\rm crit} \int_{0}^{ 2 \pi } d \theta \int_{0}^{ \RE/ \sqrt{q} } d \rho ~ q \rho \times \kappa^{\rm macro}(\rho) 
  =\\= 
  \Sigma_{\rm crit} 2 \pi \int_{0}^{ \RE/ \sqrt{q} } d \rho ~ q \rho \times  \frac{\kappa_0 \left(2 - \frac{\gamma}{2} \right)\,q^{-1/2}}{2(  r_c^2 +\rho^2)^{(\gamma-1)/2}}
  =\\
  = 
  \Sigma_{\rm crit}  \pi  q  \times \frac{\frac{1}{2}\left(2 - \frac{\gamma}{2} \right)}{(\gamma-3)/2} \frac{\kappa_0 q^{1/2}}{(  r_c^2 +\rho^2)^{(\gamma-3)/2}}
  ~\Big |_{\rho = 0}^{\rho =  \RE/ \sqrt{q}}
  =\\=
\Sigma_{\rm crit}  \pi   \times\kappa_0 \frac{\left(2 - \frac{\gamma}{2} \right)}{(\gamma-3)} \,q^{1/2}  \\ \times \Bigg[ \Big(  r_c^2 +  \RE^2/q\Big)^{(3-\gamma)/2} - \Big(  r_c^2 \Big)^{(3-\gamma)/2}
  \Bigg]
\end{multline}

The Einstein radius $\RE$ for a given set of macro-model parameters is then:

\begin{multline}
\RE^2  =  \frac{\kappa_0}{2} \frac{\left(4 - \gamma \right)}{(3-\gamma)}\,q^{1/2}  \\ \times \Bigg[ \Big(  r_c^2 +  \RE^2/q\Big)^{(3-\gamma)/2} - \Big(  r_c^2 \Big)^{(3-\gamma)/2}
  \Bigg]
\label{equ:einsteinradii}
\end{multline}

For $r_c = 0$ this definition of the Einstein radius matches the one described by \citet{vegetti2014}:

\begin{equation}
\RE^{\gamma-1} =  \frac{\kappa_0}{2} \frac{\left(4 - \gamma \right)}{(3-\gamma)}\,q^{(\gamma-2)/2}  
\label{equ:einsteinradii2}
\end{equation}

It further reduces to the definition by \citet{kormann1994} in the isothermal case where $\gamma = 2$, i.e. $\RE =  \kappa_0$.

\section{Mass function normalisation and mean number of substructures} \label{sec:mass_function_amplitude}

In this Appendix, we describe the derivation of the mass function normalisation constant from the condition that we have an average fraction of mass contained in substructures within the Einstein radius $\fsub(< \RE)$.  This condition can be written explicitly as:

\begin{multline}
\fsub(< \RE) = \Bigg <\frac{ \sum_{ \rm k=1}^{  N^{ \rm sub}} m_k } {M(< \RE)} \Bigg>_{ \P (\etasub,<\RE )   } \\ = \frac{ \mu(< \RE) <m>_{ \P(m)}}{ M(< \RE)},
\label{equ:fsub}
\end{multline}
where $\mu(< \RE)$ is the mean number of objects within the Einstein radius and $ <m>_{ \rm \P(m)}$ is the average mass of a substructure.
$\P ( N^{ \rm sub},\{m_k,\rho_k\})$ is the probability distribution associated with a substructure realisation with $N^{ \rm sub}$ clumps, each with their respective mass $m_k$ and radial position $\rho_k$.

The amplitude of the mass function $n_0$ can be determined by requiring that $\mu(< \RE)$ matches the definition given in equation (\ref{equ:fsub}), i.e.:

\begin{equation}
 \mu(< \RE)  = A(< \RE) \times  \int_{ \rm \Mmin}^{ \rm \Mmax} dm~\frac{d n^{ \rm sub}}{ dm }\,.
\end{equation}

Using the above definition for the Einstein radius, the proportionality constant in equation (\ref{equ:mf}) is: 

\begin{multline}
 n_0  = \frac{\fsub(< \RE) \times \Sigma_{ \rm crit} }{<m>_{ \rm \P(m) } \times \int_{ \rm \Mmin}^{ \rm \Mmax} dm~  m^{ \rm -\alpha}} \,.
\end{multline}

We assume that the density of substructures is constant throughout the extensions of the mock observation.
In order to obtain the mean number of objects  $ \mu $ in the considered area, we integrate the mass function over $m$ and multiply it with the area  $A^{\rm data}$ covered by the mock observation:
\begin{multline}
 \mu= A^{\rm data} \times  \int_{ \rm \Mmin}^{ \rm \Mmax} d m  \frac{d n^{ \rm sub}}{ dm }  \\= A^{\rm data} \times  \frac{\fsub(<\RE)  \times \Sigma_{ \rm crit}  }{ <m>_{ \rm \P(m) }} \,.
\end{multline}

\section{Jeffreys' prior}
\label{sec:jeffreys_prior}

For the inference task of the free parameter, $\fsub$ the  Jeffreys' prior \citep{jeffrey1946} would require us to consider the full Likelihood function $\P(\vec d|\mu)$ and marginalise it over the space data realisations. As this is computationally prohibited, we consider the Jeffreys' prior of the probability density $\P(N^{ \rm sub}|\mu)$, which being a Poisson distribution, leads to a prior of $\P(\fsub) \propto \left(\fsub\right)^{-1/2}$. We expect this simplified prior distribution to remain non-informative for our reconstruction process.
Moreover, it has the advantage of being integrable under the assumption of a finite upper limit on $\fsub$ and allows us to correctly probe a parameter which may vary by several orders of magnitude. We choose  a range of $\fsub \in [10^{-6},10^{-1}]$, which includes the case that $\mu(\fsub) + \sqrt{\mu(\fsub)}< 1$, i.e. that there are no substructures within one standard deviation of the Poisson distribution $P(N^{ \rm sub}|\mu)$.

\section{Motivating an ABC approach}\label{sec:ABC_append}

One of the main goals of this work is the reconstruction mass function parameters. The likelihood of this problem under the model assumptions specified in section \ref{sec:lens_modelling} is:
\begin{multline}
\P (\vec d | \{ f_{\rm sub} , m_{\rm hm} \} ) = \\
\sum_{N^{\rm sub}=0}^{\infty} \Bigg ( \prod_{n=1}^{N^{\rm sub}}  \int dm_n \int d \vec x_n  \P( m_n ,\vec x_n | \{ f_{\rm sub} , m_{\rm hm} \} )   \Bigg)  \times 
\\ \times \P \Big ( N \Big | \{ f_{\rm sub} , m_{\rm hm} \} \Big) 
 \times \\ \times \P \Big ( \vec d \Big | L (\etamacro, \{ \vec x_n , m_n \}_{n=1}^{N^{\rm sub}} )  \Big)
 \label{equ:full_likelihood}
\end{multline}
While we can analytically calculate the last factor of this product (corresponding to equation (\ref{equ:evidence}) once  $\mat L(\etamacro$ , $\etasub :=  \{ \vec x_n , m_n \}_{n=1}^{N^{\rm sub}})$ is given, the combination with the integrals over positions $\vec x_n$ and masses  $m_n$ as well as the summation over $N^{\rm sub}$ pose a challenge for analytical evaluation.

In order to evaluate expression (\ref{equ:full_likelihood}) one may follow a reversible jump MCMC approach as done by \citet{brewer2015} or \citet[][]{daylan2018}. However, these methods usually require to reduce the parameters in the model, for example, by switching to a parametric source rather than a free form surface brightness field. Furthermore, these approaches tend to be computationally expensive.  

Another way is to marginalize over the substructure realisations rigorously. \citet{hsueh2019} follow this approach and explicitly integrate over $\mathcal{O}(10^6)$ substructure realisations per $M_{\rm hm}$ value in their regular grid of mass function parameters. In this case, the flux-ratio analysis benefits from the small number of observables, i.e. $\dim (\vec d) \approx \mathcal{O}(10)$. The evaluation of the analogous function to equation \ref{equ:evidence} is therefore much faster than in the case of extended arcs with $\dim(\vec d) \approx \mathcal{O}(100^2)$. 
 Notice that the computational cost of expression  \ref{equ:full_likelihood} would increase even further with the inclusion of line-of-sight haloes, which constitute a possible extension of the model used in this work.

Given that our goal is to keep the physical model as general as possible, that we can generate substructure realisations with forward modelling, and that for a (somewhat) limited number of samples one will obtain conservative results with ABC, we decided to follow this approach.

\section{Importance Sampling and weighting of samples}
\label{sec:importance_acceptance}

Standard importance sampling uses weights $w_i$ and applies them to the samples that approximate the posterior distribution:

\begin{equation}
\P( \eta^{{\rm macro}} | \vec d) = \frac{1}{K^{\rm acc}} \sum_{\rm i \in {\rm acc}} \delta^D ( \etamacro - \etamacro_i) \times w_i\,,
\end{equation}

However, in our implementation, as expressed by equation (\ref{equ:abc_post}), we do not include any weight, as this is done implicitly by the acceptance strategy.
In order to justify this, we consider the following relation:
\begin{multline}
    \P( \vec x | \vec d ) = \frac{ \P(\vec d | \vec x ) }{\P (\vec d )} \P(\vec x)=  \\
    = \frac{ \P(\vec d | \vec x )  }{\P( \vec d )} \frac{ \P(\vec x) } {\Q(\vec x)} Q(\vec x)  =    \frac{\widetilde \P(\vec d| \vec x)} {\P(\vec d)}Q(\vec x)\,,
\end{multline}
with $\widetilde \P(\vec d| \vec x) = \P(\vec d| \vec x) \frac{\P(\vec x)}{Q(\vec x)}$.
This relation implies that one can interpret the importance sampling as a change in the Likelihood\footnote{Which we refer to as evidence in this paper, as it is the evidence in the case of source reconstruction.} and the prior in such a way that their changes cancel each other out. 

As we use the $P(\vec d| \vec x)$ as our distance measure, we include the weight in the distance that decides whether or not a sample gets accepted rather than multiplying the weight to the corresponding sample.
In doing so, we free ourselves from accounting for sample weights at the price of the caveat that the posterior distribution of the importance sampled parameters tend to resemble the shape of the proposal distribution.

In our case, we apply importance sampling on the macro-model parameters $\etamacro$, for which previous tests have shown that their posteriors closely resemble Gaussian distributions, therefore justifying our choice of the proposal distribution.
In order to improve the approximation, one could use Gaussian mixture models instead of a single Gaussian distribution to describe the proposition distribution. This form of the proposal distribution would allow one to account for more general shapes of posteriors.


\bsp	
\label{lastpage}

\end{document}